   \documentstyle[12pt]{article}
   \title{{\bf  A theory of tensor products for module categories for
a vertex operator algebra, II}}
    \author{Yi-Zhi Huang and James Lepowsky}
    \date{}
    
\begin{document}
    \hyphenation{Phil-a-del-phia}
    \bibliographystyle{alpha}
    \maketitle

    \input amssym.def
    \input amssym
    \newtheorem{rema}{Remark}[section]
    \newtheorem{propo}[rema]{Proposition}
    \newtheorem{theo}[rema]{Theorem}
   \newtheorem{defi}[rema]{Definition}
    \newtheorem{lemma}[rema]{Lemma}
    \newtheorem{corol}[rema]{Corollary}
     \newtheorem{exam}[rema]{Example}
	\newcommand{\ba}{\begin{array}}
	\newcommand{\ea}{\end{array}}
        \newcommand{\be}{\begin{equation}}
        \newcommand{\ee}{\end{equation}}
	\newcommand{\bea}{\begin{eqnarray}}
	\newcommand{\eea}{\end{eqnarray}}
	\newcommand{\nno}{\nonumber}
	\newcommand{\lbar}{\bigg\vert}
	\newcommand{\p}{\partial}
	\newcommand{\dps}{\displaystyle}
	\newcommand{\bra}{\langle}
	\newcommand{\ket}{\rangle}
 \newcommand{\res}{\mbox{ \rm Res}}
 \newcommand{\pf}{{\it Proof}\hspace{2ex}}
 \newcommand{\epf}{\hspace{1em}$\Box$}
 \newcommand{\epfv}{\hspace{1em}$\Box$\vspace{1em}}

\tableofcontents

\begin{abstract}
This is the second part in a series of papers presenting a theory of
tensor products for module categories for a vertex operator algebra.
In Part I (hep-th/9309076), the notions of $P(z)$- and $Q(z)$-tensor
product of two modules for a vertex operator algebra were introduced
and under a certain hypothesis, two constructions of a $Q(z)$-tensor
product were given, using certain results stated without proof.  In
Part II, the proofs of those results are supplied.
\end{abstract}

\vspace{2em}
The present paper (Part II) is a continuation of \cite{HL} (Part I), to which
the reader is referred for the necessary background, including references.
 In Part I, the notions of $P(z)$- and $Q(z)$-tensor product of two modules
for a vertex operator algebra were
introduced and under a certain hypothesis,
two constructions of a $Q(z)$-tensor product were given
based on certain results (Proposition 4.9, Proposition 5.2, Theorem 6.1
and Proposition 6.2) stated without proof. In Part II, we supply the
proofs of these results.

We fix a vertex operator algebra $V$ and $V$-modules $(W_{1}, Y_{1})$,
$(W_{2}, Y_{2})$ and $(W_{3}, Y_{3})$. We
add $y, y_{1}, y_{2}, \dots$ to our list of commuting formal
variables in Part II. The numberings of sections, formulas, etc.,
continue those of Part I.

Part II is organized as follows: Section 7 proves that the spaces of
intertwining operators of certain types are isomorphic (Proposition
4.9).  Section 8 is devoted to the proof of the commutator formula for
the vertex operators defined in Section 5 (Proposition 5.2), and a
related remark pointing out that the commutator formula and
Proposition 5.1 in fact come {}from analogous properties of an
 action on the vector space $W_{1}\otimes W_{2}$.  In Section
9, we derive some formulas involving vertex operators and the Virasoro
operators on the dual space of the tensor product vector space of two
modules. In Section 10, we prove Theorem 6.1, asserting the Jacobi
identity for the vertex operators acting on a certain subspace of this
dual space, and in Section 11, we establish Proposition 6.2, stating
the invariance of certain subspaces under these operators.

\setcounter{section}{6}
\renewcommand{\theequation}{\thesection.\arabic{equation}}
\renewcommand{\therema}{\thesection.\arabic{rema}}
\setcounter{equation}{0}
\setcounter{rema}{0}

\section{Isomorphisms between certain spaces of intertwining operators}

In this section we prove various results, including Proposition 4.9,
establishing isomorphisms between certain spaces of intertwining
operators.  These results are all simple generalizations of the
corresponding results in \cite{FHL}, where only the monodromy-free
case was considered.

Let ${\cal Y}$ be an intertwining operator of type
${W_{3}}\choose {W_{1}W_{2}}$. For any complex number $\zeta$ and any
$w_{(1)}\in W_{1}$,
${\cal Y}(w_{(1)}, y)\lbar_{y^{n}=e^{n\zeta}x^{n}, \ n\in {\Bbb C}}$ is
a well-defined element of Hom$(W_{2}, W_{3})\{ x\}$. We denote
this element by ${\cal Y}(w_{(1)}, e^{\zeta}x)$. Note that this element
depends on $\zeta$, not on just $e^{\zeta}$.
Given any $r\in {\Bbb Z}$, we define
$\Omega_{r}({\cal Y}):W_2\otimes W_1 \rightarrow  W_3\{ x\}$
by the formula
\begin{equation}
\Omega_{r}({\cal Y})(w_{(2)},x)w_{(1)} = e^{xL(-1)}
{\cal Y}(w_{(1)},e^{ (2r+1)\pi i}x)w_{(2)}.
\end{equation}
for  $w_{(1)}\in W_{1}$ and $w_{(2)}\in W_{2}$.
\begin{propo}
The operator  $\Omega_{r}({\cal Y})$  is an intertwining
operator of type  ${W_{3}}\choose {W_{2} \ W_{1}}$.  Moreover,
\begin{equation}
\Omega_{-r-1}(\Omega_{r}({\cal Y}))=\Omega_{r}(\Omega_{-r-1}({\cal Y}))
 = {\cal Y}.
\end{equation}
In particular, the correspondence  ${\cal Y} \mapsto
\Omega_{r}({\cal Y})$  defines a linear isomorphism {from}
${\cal V}^{W_{3}}_{W_{1}W_{2}}$ to
${\cal V}^{W_{3}}_{W_{2}W_{1}}$,  and we have
\begin{equation}
N^{W_{3}}_{W_{1}W_{2}} = N^{W_{3}}_{W_{2}W_{1}}.
\end{equation}
\end{propo}
\pf
The proof of this result is a straightforward generalization of  the proof
of Proposition 5.4.7 in \cite{FHL}. The lower truncation condition (2.44)
is clear. {From} the Jacobi identity
\begin{eqnarray}
\lefteqn{\dps x^{-1}_0\delta \left( {x_1-y\over x_0}\right)
Y_3(v,x_1){\cal Y}(w_{(1)},y)w_{(2)}}\nno\\
&&\hspace{2em}- x^{-1}_0\delta \left( {y-x_1\over -x_0}\right)
{\cal Y}(w_{(1)},y)Y_2(v,x_1)w_{(2)}\nno \\
&&{\dps = y^{-1}\delta \left( {x_1-x_0\over y}\right)
{\cal Y}(Y_1(v,x_0)w_{(1)},y)
w_{(2)}}
\end{eqnarray}
for ${\cal Y}$, we obtain
\begin{eqnarray}
\dps x^{-1}_0\delta \left( {x_1-y\over x_0}\right)
e^{-yL(-1)}Y_3(v,x_1){\cal Y}(w_{(1)},y)w_{(2)}
\lbar_{y^{n}=e^{n(2r+1)\pi i}x_{2}^{n}, \; n\in {\Bbb C}}\hspace{1.5em}\nno\\
- x^{-1}_0\delta \left( {y-x_1\over -x_0}\right)
e^{-yL(-1)}{\cal Y}(w_{(1)},y)Y_2(v,x_1)w_{(2)}
\lbar_{y^{n}=e^{n (2r+1)\pi i}x_{2}^{n}, \; n\in {\Bbb C}}\nno \\
{\dps = y^{-1}\delta \left( {x_1-x_0\over y}\right)
e^{-yL(-1)}{\cal Y}(Y_1(v,x_0)w_{(1)},y)
w_{(2)}\lbar_{y^{n}=e^{n (2r+1)\pi i}x_{2}^{n}, \; n\in {\Bbb C}}}.
\end{eqnarray}
The first term of the left-hand side of (7.5) is equal to
\begin{eqnarray*}
&{\dps x^{-1}_0\delta \left( {x_1-y\over x_0}\right)
Y_3(v,x_1-y)e^{-yL(-1)}{\cal Y}(w_{(1)},y)w_{(2)}
\lbar_{y^{n}=e^{n (2r+1)\pi i}x_{2}^{n}, \; n\in {\Bbb C}}}&\nno\\
&{\dps =x^{-1}_0\delta \left( {x_1+x_{2}\over x_0}\right)
Y_3(v,x_0)\Omega_{r}({\cal Y})(w_{(2)},x_{2})w_{(1)},}&
\end{eqnarray*}
the second term  is equal to
$$
- x^{-1}_0\delta \left( {-x_{2}-x_1\over -x_0}\right)
\Omega_{r}({\cal Y})(Y_2(v,x_1)w_{(2)},x_{2})w_{(1)}
$$
and the right-hand side of (7.5) is equal to
$$
-x_{2}^{-1}\delta \left( {x_1-x_0\over -x_{2}}\right)
\Omega_{r}({\cal Y})(w_{(2)},x_{2})Y_1(v,x_0)w_{(1)}.
$$
Substituting  into (7.5) we obtain
\begin{eqnarray}
\lefteqn{x^{-1}_0\delta \left( {x_1+x_{2}\over x_0}\right)
Y_3(v,x_0)\Omega_{r}({\cal Y})(w_{(2)},x_{2})w_{(1)}}\nno\\
&&\hspace{2em}- x^{-1}_0\delta \left( {x_{2}+x_1\over x_0}\right)
\Omega_{r}({\cal Y})(Y_2(v,x_1)w_{(2)},x_{2})w_{(1)}\nno\\
&&=-x_{2}^{-1}\delta \left( {x_1-x_0\over -x_{2}}\right)
\Omega_{r}({\cal Y})(w_{(2)},x_{2})Y_1(v,x_0)w_{(1)},
\end{eqnarray}
which by (2.6) is equivalent to
\begin{eqnarray}
\lefteqn{x^{-1}_1\delta \left( {x_0-x_{2}\over x_1}\right)
Y_3(v,x_0)\Omega_{r}({\cal Y})(w_{(2)},x_{2})w_{(1)}}\nno\\
&&\hspace{2em}-x_{1}^{-1}\delta \left( {x_2-x_0\over -x_{1}}\right)
\Omega_{r}({\cal Y})(w_{(2)},x_{2})Y_1(v,x_0)w_{(1)}\nno\\
&&= x^{-1}_2\delta \left( {x_{0}-x_1\over x_2}\right)
\Omega_{r}({\cal Y})(Y_2(v,x_1)w_{(2)},x_{2})w_{(1)}.
\end{eqnarray}
This in turn is the Jacobi identity for $\Omega_{r}({\cal Y})$. The
$L(-1)$-derivative property of $\Omega_{r}({\cal Y})$
is easily proved as in the proof of Proposition 5.4.7 of \cite{FHL}.
The identity (7.2) is obvious {from} the definitions of
$\Omega_{r}({\cal Y})$ and $\Omega_{-r-1}({\cal Y})$. \epf

\begin{rema}
{\rm As  Propositions 4.7 and 7.1 suggest, for each triple $s_{1},
s_{2}, s_{3}\in {\Bbb Z}$ the
intertwining operator ${\cal Y}$ gives rise to an intertwining operator
${\cal Y}_{[s_{1}, s_{2}. s_{3}]}$ of the same type, defined by
\begin{equation}
{\cal Y}_{[s_{1}, s_{2}, s_{3}]}(w_{(1)}, x)=e^{2s_{1}\pi iL(0)}
{\cal Y}(e^{2s_{2}\pi
iL(0)}w_{(1)}, x)e^{2s_{3}\pi iL(0)}
\end{equation}
for $w_{(1)}\in W_{1}$; it is easy to see that ${\cal Y}_{[s_{1}, s_{2},
s_{3}]}$ is again
an intertwining operator. We have ${\cal Y}_{[0, 0, 0]}={\cal Y}$ and for
$r_{1}, r_{2}, r_{3}, s_{1}, s_{2}, s_{3}\in {\Bbb Z}$,
\begin{equation}
({\cal Y}_{[r_{1}, r_{2}, r_{3}]})_{[s_{1}, s_{2}, s_{3}]}
={\cal Y}_{[r_{1}+s_{1}, r_{2}+s_{2}, r_{3}+s_{3}]}.
\end{equation}
For any $a\in {\Bbb C}$, we have the formula
\begin{equation}
e^{aL(0)}{\cal Y}(w_{(1)},x)e^{-aL(0)}={\cal Y}(e^{a L(0)}w_{(1)}, e^{a}x)
\end{equation}
whose proof is similar to that of formula (5.4.22) of \cite{FHL}.
{}From this formula,  we see that  (7.2) generalizes to:
\begin{eqnarray}
\Omega_{s}(\Omega_{r}({\cal Y}))&=&{\cal Y}_{[r+s+1, -(r+s+1), -(r+s+1)]}\nno\\
&=&{\cal Y}(\cdot, e^{2(r+s+1)\pi i}\cdot).
\end{eqnarray}}
\end{rema}

Given an intertwining operator ${\cal Y}$ of type ${W_{3}}\choose
{W_{1}W_{2}}$ and an integer $r\in {\Bbb Z}$, we define the {\it
$r$-contragredient operator of ${\cal Y}$} to be the linear map
\begin{eqnarray}
W_1 \otimes  W'_3& \rightarrow&   W'_2\{z\}\nno\\
w_{(1)}\otimes w'_{(3)} &\mapsto  &A_{r}({\cal Y})(w_{(1)},x)w'_{(3)}
\end{eqnarray}
given by
\begin{eqnarray}
\lefteqn{\langle A_{r}({\cal Y})(w_{(1)},x)w'_{(3)},w_{(2)}\rangle _{W_2}=}
\nno\\
&&= \langle w'_{(3)}, {\cal Y}(e^{xL(1)}e^{(2r+1)\pi i L(0)}
x^{-2L(0)}w_{(1)},x^{-1})w_{(2)}
\rangle _{W_3},
\end{eqnarray}
where $w_{1}\in W_{1}$, $w_{2}\in W_{2}$, $w'_{(3)}\in W'_{(3)}$ and
$\langle \cdot ,\cdot \rangle _{W_2}$ and $\langle \cdot ,\cdot
\rangle _{W_3}$ are the pairings for $W'_2,W_2$ and for $W'_3,W_3$,
respectively. Note that for the case $W_{1}=V$ and $W_{2}=W_{3}=W$,
the operator $A_{r}({\cal Y})$ agrees with the contragredient vertex
operator ${\cal Y}'$ (recall (2.48), (2.49)) for any $r\in {\Bbb Z}$.

\begin{propo}
The $r$-contragredient operator $A_{r}({\cal Y})$ of an intertwining operator
${\cal Y}$ of type ${W_3}\choose {W_1 W_2}$ is an
intertwining operator of type  ${W'_2}\choose {W_1 W'_3}$.
Moreover,
\begin{equation}
A_{-r-1}(A_{r}({\cal Y}))=A_{r}(A_{-r-1}({\cal Y}))={\cal Y}.
\end{equation}
In particular, the correspondence
${\cal Y} \mapsto  A_{r}({\cal Y})$ defines a linear isomorphism {from}
${\cal V}^{W_{3}}_{W_{1}W_{2}}$ to  ${\cal V}^{W'_{2}}_{W_{1}W'_{3}}$,
and we have
\begin{equation}
N^{W_{3}}_{W_{1}W_{2}}= N^{W'_{2}}_{W_{1}W'_{3}}.
\end{equation}
\end{propo}
 \pf
The proof  of this result is a straightforward generalization of the proof
of Theorem 5.5.1 and Proposition 5.5.2 in \cite{FHL}. The lower
truncation condition (2.44) being easy to verify, we want to prove the identity
\begin{eqnarray}
\lefteqn{\langle x^{-1}_0\delta \left( {x_1-x_2\over x_0}\right)
Y_{2}'(v,x_1)A_{r}({\cal Y})(w_{(1)},x_2)w'_{(3)},w_{(2)}\rangle_{W_2}}\nno \\
&&\;\;\;\;- \langle x^{-1}_0\delta \left( {x_2-x_1\over -x_0}\right)
A_{r}({\cal Y})(w_{(1)},x_2)Y_{3}'(v,x_1)w'_{(3)},w_{(2)}\rangle_{W_2}\nno\\
&&=\langle x^{-1}_2\delta \left( {x_1-x_0\over x_2}\right)
A_{r}({\cal Y})(Y_{1}(v,x_0)w_{(1)},x_2)w'_{(3)},w_{(2)}\rangle_{W_2}.
\end{eqnarray}
But by the definitions (2.49) and (7.13) we have
\begin{eqnarray}
\lefteqn{\langle Y_{2}'(v,x_1)A_{r}({\cal Y})
(w_{(1)},x_2)w'_{(3)},w_{(2)}\rangle_{W_2}=}\nno  \\
&&= \langle w'_{(3)},{\cal Y}(e^{x_2L(1)}e^{(2r+1)\pi iL(0)}
x_2^{-2L(0)}w_{(1)},x^{-1}_2)\cdot \nno\\
&&\hspace{8em}\cdot Y_{2}(e^{x_1L(1)}(-x^{-2}_1)^{L(0)}v,x^{-1}_1)w_{(2)}
\rangle_{W_3},
\end{eqnarray}
\begin{eqnarray}
\lefteqn{\langle A_{r}({\cal Y})(w_{(1)},x_2)
Y_{3}'(v,x_1)w'_{(3)},w_{(2)}\rangle_{W_2}=}\nno  \\
&&=\langle w'_{(3)},Y_{3}(e^{x_1L(1)}(-x^{-2}_1)^{L(0)}v,x^{-1}_1)\cdot \nno\\
&&\hspace{6em}\cdot {\cal Y}(e^{x_2L(1)}e^{(2r+1)\pi iL(0)}
x_2^{-2L(0)}w_{(1)},x^{-1}_2)w_{(2)}\rangle_{W_3},
\end{eqnarray}
\begin{eqnarray}
\lefteqn{\langle A_{r}({\cal Y})
(Y_{1}(v,x_0)w_{(1)},x_2)w'_{(3)},w_{(2)}\rangle_{W_2}= }\nno \\
&&= \langle w'_{(3)},{\cal Y}(e^{x_2L(1)}e^{(2r+1)\pi iL(0)}
x_2^{-2L(0)}
Y_{1}(v,x_0)w_{(1)},x^{-1}_2)w_{(2)}\rangle_{W_3},
\end{eqnarray}
and {from} the Jacobi identity for  ${\cal Y}$  we have
\begin{eqnarray}
\lefteqn{\langle w'_{(3)},\left( {-x_0\over x_1x_2}\right) ^{-1}\delta
 \left( {x^{-1}_1-x^{-1}_2\over -x_0/x_1x_2} \right)
Y_{3}(e^{x_1L(1)}(-x^{-2}_1)^{L(0)}v,x^{-1}_1)\cdot}\nno \\
&&\;\;\;\;\;\;\;\;\;\;\cdot {\cal Y}(e^{x_2L(1)}e^{(2r+1)\pi iL(0)}
x_2^{-2L(0)}w_{(1)},x^{-1}_2)w_{(2)}\rangle_{W_3}\nno \\
&&- \langle w'_{(3)},\left( {-x_0\over x_1x_2}\right) ^{-1}\delta
\left( {x^{-1}_2-x^{-1}_1\over x_0/x_1x_2}\right)
{\cal Y}(e^{x_2L(1)}e^{(2r+1)\pi iL(0)}
x_2^{-2L(0)}w_{(1)},x^{-1}_2)\cdot \nno\\
&&\;\;\;\;\;\;\;\;\;\;\cdot
 Y_{2}(e^{x_1L(1)}(-x^{-2}_1)^{L(0)}v,x^{-1}_1)w_{(2)}\rangle_{W_3} \nno\\
&&= \langle w'_{(3)},(x^{-1}_2)^{-1}
\delta \left( {x^{-1}_1+x_0/x_1x_2\over x^{-1}_2}\right)
{\cal Y}(Y_{1}(e^{x_1L(1)}(-x^{-2}_1)^{L(0)}v,
-x_0/x_1x_2)\cdot \nno\\
&&\;\;\;\;\;\;\;\;\;\;\cdot e^{x_2L(1)}e^{(2r+1)\pi iL(0)}
x_2^{-2L(0)}w_{(1)},x^{-1}_2)w_{(2)}\rangle_{W_3},
\end{eqnarray}
or equivalently,
\begin{eqnarray}
\lefteqn{- \langle w'_{(3)},x^{-1}_0\delta \left( {x_2-x_1\over -x_0}\right)
Y_{3}(e^{x_1L(1)}(-x^{-2}_1)^{L(0)}v,x^{-1}_1)\cdot }\nno\\
&&\;\;\;\;\;\;\;\;\;\;\cdot {\cal Y}(e^{x_2L(1)}e^{(2r+1)\pi iL(0)}
x_2^{-2L(0)}w_{(1)},
x^{-1}_2)w_{(2)}\rangle_{W_3} \nno\\
&&\;\;\;\;+ \langle w'_{(3)},x^{-1}_0
\delta \left( {x_1-x_2\over x_0}\right)
{\cal Y}(e^{x_2L(1)}e^{(2r+1)\pi
iL(0)}x_2^{-2L(0)}w_{(1)},x^{-1}_2)\cdot \nno\\
&&\;\;\;\;\;\;\;\;\;\;\cdot Y_{2}(e^{x_1L(1)}(-x^{-2}_1)^{L(0)}v,
x^{-1}_1)w_{(2)}\rangle_{W_3} \nno\\ &&= \langle w'_{(3)},x^{-1}_1\delta
\left( {x_2+x_0\over x_{1}}\right) {\cal Y}(Y_{1}
(e^{x_1L(1)}(-x^{-2}_1)^{L(0)}v,-x_0/x_1x_2)\cdot\nno\\
&&\;\;\;\;\;\;\;\;\;\;\cdot e^{x_2L(1)}e^{(2r+1)\pi iL(0)}
x_2^{-2L(0)}w_{(1)},x^{-1}_2)w_{(2)}\rangle_{W_3}.
\end{eqnarray}
Substituting (7.17), (7.18) and (7.19) into (7.16) and then comparing with
(7.21), we see
that the proof of (7.16) is reduced to the proof of
\begin{eqnarray}
\lefteqn{x^{-1}_1\delta \left( {x_2+x_0\over x_1}\right)
{\cal Y}(e^{x_2L(1)}e^{(2r+1)\pi iL(0)}
x_2^{-2L(0)}Y_{1}(v,x_0)w_{(1)},x^{-1}_2)}\nno \\
&&= x^{-1}_1\delta \left( {x_2+x_0\over x_1}\right)
 {\cal Y}(Y_{1}(e^{x_1L(1)}(-x^{-2}_1)^{L(0)}v,-x_0/x_1x_2)\cdot \nno\\
&&\;\;\;\;\;\;\;\;\;\;\cdot e^{x_2L(1)}e^{(2r+1)\pi iL(0)}
x_2^{-2L(0)}w_{(1)},x^{-1}_2),
\end{eqnarray}
or of
\begin{eqnarray}
\lefteqn{{\cal Y}(e^{x_2L(1)}e^{(2r+1)\pi iL(0)}
x_2^{-2L(0)}Y_{1}(v,x_0)w_{(1)},x^{-1}_2)=}\nno \\
&&= {\cal Y}(Y_{1}
(e^{(x_2+x_0)L(1)}(-(x_2+x_0)^{-2})^{L(0)}v,-x_0/(x_2+x_0)x_2)\cdot\nno \\
&&\;\;\;\;\;\;\;\;\;\;\cdot e^{x_2L(1)}e^{(2r+1)\pi iL(0)}
x_2^{-2L(0)}w_{(1)},x^{-1}_2).
\end{eqnarray}
We see that we need only prove
\begin{eqnarray}
\lefteqn{e^{x_2L(1)}e^{(2r+1)\pi iL(0)}
x_2^{-2L(0)}Y_{1}(v,x_0)=}\nno\\
&&=Y_{1}(e^{(x_2+x_0)L(1)}(-(x_2+x_0)^{-2})^{L(0)}v,
-x_0/(x_2+x_0)x_2)\cdot\nno\\
&&\;\;\;\;\;\;\;\;\;\;\cdot e^{x_2L(1)}e^{(2r+1)\pi iL(0)}
x_2^{-2L(0)}
\end{eqnarray}
or equivalently, the conjugation formula
\begin{eqnarray}
\lefteqn{e^{x_{2}L(1)}e^{(2r+1)\pi iL(0)}
x_{2}^{-2L(0)}Y_{1}(v,x_0)x_{2}^{2L(0)}e^{-(2r+1)\pi iL(0)}
e^{-x_{2}L(1)}}\nno \\
&&= Y_1(e^{(x_{2}+x_0)L(1)}(-(x_{2}+x_0)^{-2})^{L(0)}v,-x_0/(x_{2}+x_0)x_{2})
\end{eqnarray}
for $v\in V$, acting on the module $W_{1}$. But
formula (7.25) follows {from} Lemma 5.2.3 of \cite{FHL} and the
 formula
\begin{equation}
e^{(2r+1)\pi iL(0)}Y_{1}(v,x)e^{-(2r+1)\pi iL(0)}=Y_{1}((-1)^{L(0)}v, -x),
\end{equation}
which is a special case of (7.10).
This establishes the Jacobi identity, and the $L(-1)$-derivative
property follows {}from the same argument used in the proof of
Theorem 5.5.1 of \cite{FHL}. Finally the relation (7.14) is obtained by
combining the argument in the proof of Proposition 5.3.1 of \cite{FHL} and the
formula
\begin{equation}
e^{(2r+1)\pi iL(0)}x^{2L(0)}e^{xL(1)}x^{-2L(0)}e^{-(2r+1)\pi
iL(0)}=e^{-x^{-1}L(1)}
\end{equation}
on $W_{1}$, whose proof is similar to that of formula (5.3.1) of \cite{FHL}.
\epf

\begin{rema}
{\rm This argument shows that for any $r, s\in {\Bbb Z}$, formula
(7.14) generalizes to:
\begin{equation}
A_{s}(A_{r}({\cal Y}))={\cal Y}_{[0, r+s+1, 0]}.
\end{equation}}
\end{rema}

{From} Proposition 7.1,  for any integer $r_{1}$ we have an
 isomorphism $\Omega_{r_{1}}$ {from} ${\cal
V}^{W_{3}}_{W_{1}W_{2}}$ to ${\cal V}^{W_{3}}_{W_{2}W_{1}}$, and {from}
Proposition 7.3,  for any integer $r_{2}$ we have an
 isomorphism $A_{r_{2}}$ {from} ${\cal V}^{W_{3}}_{W_{2}W_{1}}$
to ${\cal V}^{W'_{1}}_{W_{2}W'_{3}}$. By Proposition 7.1 again,
 for any integer $r_{3}$ there is an  isomorphism, which we again denote
$\Omega_{r_{3}}$, {from} ${\cal V}^{W'_{1}}_{W_{2}W'_{3}}$ to ${\cal
V}^{W'_{1}}_{W'_{3}W_{2}}$. Thus for any triple $(r_{1}, r_{2},
r_{3})$ of integers, we have an isomorphism $\Omega_{r_{3}}\circ A_{r_{2}}\circ
\Omega_{r_{1}}$ {from} ${\cal V}^{W_{3}}_{W_{1}W_{2}}$ to ${\cal
V}^{W'_{1}}_{W'_{3}W_{2}}$.  Let ${\cal Y}$ be an intertwining
operator in ${\cal V}^{W_{3}}_{W_{1}W_{2}}$ and $w_{(1)}$, $w_{(2)}$,
$w'_{(3)}$ elements of $W_{1}$, $W_{2}$, $W'_{3}$, respectively.
{From} the definitions of $\Omega_{r_{1}}$, $A_{r_{2}}$ and
$\Omega_{r_{3}}$, we have
\begin{eqnarray}
\lefteqn{\langle (\Omega_{r_{3}}\circ A_{r_{2}}\circ
\Omega_{r_{1}})({\cal Y})(w'_{(3)}, x)w_{(2)}, w_{(1)}\rangle_{W_1}=}\nno\\
&&=\langle e^{xL(-1)}A_{r_{2}}(\Omega_{r_{1}}
({\cal Y}))(w_{(2)}, e^{(2r_{3}+1)\pi i}x)w'_{(3)},
w_{(1)}\rangle_{W_1}\nno\\
&&=\langle A_{r_{2}}(\Omega_{r_{1}}
({\cal Y}))(w_{(2)}, e^{(2r_{3}+1)\pi i}x)w'_{(3)},
e^{xL(1)}w_{(1)}\rangle_{W_1}\nno\\
&&=\langle w'_{(3)}, \Omega_{r_{1}}({\cal Y})(e^{-xL(1)}e^{(2r_{2}+1)\pi iL(0)}
e^{-2(2r_{3}+1)\pi iL(0)}x^{-2L(0)}w_{(2)}, \nno\\
&&\hspace{10em}e^{-(2r_{3}+1)\pi i}x^{-1})
e^{xL(1)}w_{(1)}\rangle_{W_3}
\nno\\
&&=\langle w'_{(3)}, e^{-x^{-1}L(-1)}{\cal Y}(e^{xL(1)}w_{(1)},
e^{(2r_{1}+1)\pi i}e^{-(2r_{3}+1)\pi i}x^{-1})\cdot \nno\\
&&\hspace{4em}\cdot e^{-xL(1)}e^{(2r_{2}+1)\pi iL(0)}
e^{-2(2r_{3}+1)\pi iL(0)}x^{-2L(0)}w_{(2)}\rangle_{W_3}
\nno\\
&&=\langle e^{-x^{-1}L(1)}w'_{(3)}, {\cal Y}(e^{xL(1)}w_{(1)},
e^{2(r_{1}-r_{3})\pi i}x^{-1})\cdot \nno\\
&&\hspace{6em}\cdot e^{-xL(1)}e^{(2(r_{2}-2r_{3}-1)+1)\pi
iL(0)}x^{-2L(0)}w_{(2)}\rangle_{W_3}.
\end{eqnarray}
{From} (7.29) we see that $\Omega_{r_{3}}\circ A_{r_{2}}\circ
\Omega_{r_{1}}$ depends only on $r_{2}-2r_{3}-1$ and $r_{1}-r_{3}$, and
the operators $\Omega_{r_{3}}\circ A_{r_{2}}\circ \Omega_{r_{1}}$ with
different
$r_{1}-r_{3}$ but the same $r_{2}-2r_{3}-1$ differ {from} each other
only by automorphisms of ${\cal V}^{W_{3}}_{W_{1}W_{2}}$. Thus
for our purpose, we need only consider those isomorphisms such that
$r_{1}-r_{3}=0$.  Given any integer $r$, we choose two integers $r_{2}$
and $r_{3}$ such that $r=r_{2}-2r_{3}-1$ and we define
\begin{equation}
B_{r}=\Omega_{r_{3}}\circ A_{r_{2}}\circ \Omega_{r_{3}}.
\end{equation}
{From} (7.29) we see that $B_{r}$ is independent of the choices of
$r_{2}$ and $r_{3}$ and that
(4.31) holds.
This proves Proposition 4.9.

Set
\begin{equation}
N_{W_{1}W_{2}W_{3}}=N^{W'_{3}}_{W_{1}W_{2}}.
\end{equation}
Then {from} Proposition 7.1 we have
\begin{equation}
N_{W_{1}W_{2}W_{3}}=N_{W_{2}W_{1}W_{3}}
\end{equation}
and {from} Proposition 7.3 we have
\begin{equation}
N_{W_{1}W_{2}W_{3}}=N_{W_{1}W_{3}W_{2}},
\end{equation}
so that we obtain (cf. formula (5.5.8) of \cite{FHL}):
\begin{propo}
For any permutation $\sigma$ of $\{1,2,3\}$,
\begin{equation}
N_{W_{1}W_{2}W_{3}}=N_{W_{\sigma(1)}W_{\sigma(2)}W_{\sigma(3)}}.
\hspace{3em}\Box
\end{equation}
\end{propo}

\renewcommand{\theequation}{\thesection.\arabic{equation}}
\renewcommand{\therema}{\thesection.\arabic{rema}}
\setcounter{equation}{0}
\setcounter{rema}{0}

\section{Commutator formula for  vertex operators on $(W_{1}\otimes
W_{2})^{*}$}

We prove Proposition 5.2 in this section. The reader should note the
well-definedness of each expression and the justifiability of each use of
a $\delta$-function property in the argument which follows.
Let $\lambda
\in (W_{1}\otimes W_{2})^{*}$, $v_{1}, v_{2}\in V$, $w_{(1)}\in W_{1}$
and $w_{(2)}\in W_{2}$. By (5.4),
\bea
\lefteqn{(Y'_{Q(z)}(v_{1}, x_{1})Y'_{Q(z)}(v_{2}, x_{2})
\lambda)(w_{(1)}\otimes w_{(2)})=}\nno\\
&&=\mbox{\rm Res}_{y_{1}}x_{1}^{-1}\delta
\left(\frac{y_{1}-z}{x_{1}}\right)(Y'_{Q(z)}(v_{2}, x_{2})
\lambda)(Y_{1}^{*}(v_{1}, y_{1})w_{(1)}\otimes w_{(2)})\nno\\
&&\quad -\mbox{\rm Res}_{y_{1}}x_{1}^{-1}\delta
\left(\frac{z-y_{1}}{-x_{1}}\right)(Y'_{Q(z)}(v_{2}, x_{2})
\lambda)(w_{(1)}\otimes Y_{2}(v_{1}, y_{1})w_{(2)})\nno\\
&&=\mbox{\rm Res}_{y_{1}}\mbox{\rm Res}_{y_{2}}x_{1}^{-1}\delta
\left(\frac{y_{1}-z}{x_{1}}\right)x_{2}^{-1}\delta
\left(\frac{y_{2}-z}{x_{2}}\right)\cdot \nno\\
&&\hspace{4em}\cdot \lambda(Y_{1}^{*}(v_{2}, y_{2})Y_{1}^{*}
(v_{1}, y_{1})w_{(1)}\otimes w_{(2)})
\nno\\
&&\quad -\mbox{\rm Res}_{y_{1}}\mbox{\rm Res}_{y_{2}}x_{1}^{-1}\delta
\left(\frac{y_{1}-z}{x_{1}}\right)x_{2}^{-1}\delta
\left(\frac{z-y_{2}}{-x_{2}}\right)\cdot \nno\\
&&\hspace{4em}\cdot \lambda(Y_{1}^{*}(v_{1}, y_{1})w_{(1)}
\otimes Y_{2}(v_{2}, y_{2})w_{(2)})\nno\\
&&\quad -\mbox{\rm Res}_{y_{1}}\mbox{\rm Res}_{y_{2}}x_{1}^{-1}\delta
\left(\frac{z-y_{1}}{-x_{1}}\right)x_{2}^{-1}\delta
\left(\frac{y_{2}-z}{x_{2}}\right)\cdot \nno\\
&&\hspace{4em} \cdot \lambda(Y_{1}^{*}(v_{2}, y_{2})w_{(1)}
\otimes Y_{2}(v_{1}, y_{1})w_{(2)})\nno\\
&&\quad +\mbox{\rm Res}_{y_{1}}\mbox{\rm Res}_{y_{2}}x_{1}^{-1}\delta
\left(\frac{z-y_{1}}{-x_{1}}\right)x_{2}^{-1}\delta
\left(\frac{z-y_{2}}{-x_{2}}\right)\cdot \nno\\
&&\hspace{4em}\cdot \lambda(w_{(1)}
\otimes Y_{2}(v_{2}, y_{2})Y_{2}(v_{1}, y_{1})w_{(2)}).
\eea
Transposing the subscripts $1$ and $2$ of the symbols $v$, $x$ and $y$,
we have
\bea
\lefteqn{(Y'_{Q(z)}(v_{2}, x_{2})Y'_{Q(z)}(v_{1}, x_{1})\lambda)(w_{(1)}
\otimes w_{(2)})=}\nno\\
&&=\mbox{\rm Res}_{y_{2}}\mbox{\rm Res}_{y_{1}}x_{2}^{-1}
\delta\left(\frac{y_{2}-z}{x_{2}}\right)x_{1}^{-1}
\delta\left(\frac{y_{1}-z}{x_{1}}\right)\cdot \nno\\
&&\hspace{4em}\cdot \lambda(Y_{1}^{*}
(v_{1}, y_{1})Y_{1}^{*}(v_{2}, y_{2})w_{(1)}\otimes w_{(2)})\nno\\
&&\quad -\mbox{\rm Res}_{y_{2}}\mbox{\rm Res}_{y_{1}}x_{2}^{-1}
\delta\left(\frac{y_{2}-z}{x_{2}}\right)x_{1}^{-1}
\delta\left(\frac{z-y_{1}}{-x_{1}}\right)\cdot \nno\\
&&\hspace{4em}\cdot \lambda(Y_{1}^{*}(v_{2}, y_{2})w_{(1)}
\otimes Y_{2}(v_{1}, y_{1})w_{(2)})\nno\\
&&\quad -\mbox{\rm Res}_{y_{2}}\mbox{\rm Res}_{y_{1}}x_{2}^{-1}
\delta\left(\frac{z-y_{2}}{-x_{2}}\right)x_{1}^{-1}\delta
\left(\frac{y_{1}-z}{x_{1}}\right)\cdot \nno\\
&&\hspace{4em}\cdot \lambda(Y_{1}^{*}(v_{1}, y_{1})w_{(1)}
\otimes Y_{2}(v_{2}, y_{2})w_{(2)})\nno\\
&&\quad +\mbox{\rm Res}_{y_{2}}\mbox{\rm Res}_{y_{1}}x_{2}^{-1}\delta
\left(\frac{z-y_{2}}{-x_{2}}\right)x_{1}^{-1}\delta\left
(\frac{z-y_{1}}{-x_{1}}\right)\cdot \nno\\
&&\hspace{4em}\cdot \lambda(w_{(1)}\otimes Y_{2}(v_{1}, y_{1})
Y_{2}(v_{2}, y_{2})w_{(2)}).
\eea
The equalities (8.1) and (8.2) give
\bea
\lefteqn{([Y'_{Q(z)}(v_{1}, x_{1}), Y'_{Q(z)}(v_{2}, x_{2})]\lambda)
(w_{(1)}\otimes w_{(2)})=}\nno\\
&&=\mbox{\rm Res}_{y_{2}}\mbox{\rm Res}_{y_{1}}x_{1}^{-1}
\delta\left(\frac{y_{1}-z}{x_{1}}\right)x_{2}^{-1}\delta
\left(\frac{y_{2}-z}{x_{2}}\right)\cdot \nno\\
&&\hspace{4em}\cdot \lambda([Y_{1}^{*}(v_{2}, y_{2}),
Y_{1}^{*}(v_{1}, y_{1})]w_{(1)}
\otimes w_{(2)})\nno\\
&&\quad -\mbox{\rm Res}_{y_{2}}\mbox{\rm Res}_{y_{1}}x_{1}^{-1}
\delta\left(\frac{z-y_{1}}{-x_{1}}\right)x_{2}^{-1}
\delta\left(\frac{z-y_{2}}{-x_{2}}\right)\cdot \nno\\
&&\hspace{4em}\cdot \lambda(w_{(1)}\otimes [Y_{2}(v_{1}, y_{1}),
Y_{2}(v_{2}, y_{2})]
w_{(2)})\nno\\
&&=\mbox{\rm Res}_{y_{2}}\mbox{\rm Res}_{y_{1}}x_{1}^{-1}
\delta\left(\frac{y_{1}-z}{x_{1}}\right)x_{2}^{-1}
\delta\left(\frac{y_{2}-z}{x_{2}}\right)\cdot \nno\\
&&\hspace{4em}\cdot \lambda\left(\mbox{\rm Res}_{x_{0}}y_{2}^{-1}
\delta\left(\frac{y_{1}-x_{0}}{y_{2}}\right)
Y^{*}_{1}(Y(v_{1}, x_{0})v_{2}, y_{2})w_{(1)}\otimes w_{(2)}\right) \nno\\
&&\quad -\mbox{\rm Res}_{y_{2}}\mbox{\rm Res}_{y_{1}}x_{1}^{-1}
\delta\left(\frac{z-y_{1}}{-x_{1}}\right)x_{2}^{-1}
\delta\left(\frac{z-y_{2}}{-x_{2}}\right)\cdot \nno\\
&&\hspace{4em}\cdot \lambda\left(
w_{(1)}\otimes \mbox{\rm Res}_{x_{0}}y_{2}^{-1}
\delta\left(\frac{y_{1}-x_{0}}{y_{2}}\right)Y_{2}(Y(v_{1}, x_{0})v_{2},
y_{2})w_{(2)}\right)\nno\\
&&=\mbox{\rm Res}_{x_{0}}\mbox{\rm Res}_{y_{2}}
\mbox{\rm Res}_{y_{1}}x_{1}^{-1}\delta\left(\frac{y_{1}-z}{x_{1}}\right)
x_{2}^{-1}\delta\left(\frac{y_{2}-z}{x_{2}}\right)y_{2}^{-1}
\delta\left(\frac{y_{1}-x_{0}}{y_{2}}\right)\cdot \nno\\
&&\hspace{4em}\cdot \lambda(
Y^{*}_{1}(Y(v_{1}, x_{0})v_{2}, y_{2})w_{(1)}\otimes w_{(2)})\nno\\
&&\quad -\mbox{\rm Res}_{x_{0}}\mbox{\rm Res}_{y_{2}}\mbox{\rm Res}_{y_{1}}
x_{1}^{-1}\delta\left(\frac{z-y_{1}}{-x_{1}}\right)x_{2}^{-1}
\delta\left(\frac{z-y_{2}}{-x_{2}}\right)y_{2}^{-1}\delta
\left(\frac{y_{1}-x_{0}}{y_{2}}\right)\cdot \nno\\
&&\hspace{4em}\cdot \lambda(w_{(1)}\otimes Y_{2}(Y(v_{1}, x_{0})v_{2},
y_{2})w_{(2)}).
\eea
Since
\bea
\lefteqn{x_{1}^{-1}\delta\left(\frac{y_{1}-z}{x_{1}}\right)x_{2}^{-1}
\delta\left(\frac{y_{2}-z}{x_{2}}\right)y_{2}^{-1}
\delta\left(\frac{y_{1}-x_{0}}{y_{2}}\right)=}\nno\\
&&=y_{1}^{-1}\delta\left(\frac{x_{1}+z}{y_{1}}\right)y_{2}^{-1}
\delta\left(\frac{x_{2}+z}{y_{2}}\right)(x_{2}+z)^{-1}
\delta\left(\frac{(x_{1}+z)-x_{0}}{x_{2}+z}\right)\nno\\
&&=y_{1}^{-1}\delta\left(\frac{x_{1}+z}{y_{1}}\right)y_{2}^{-1}
\delta\left(\frac{x_{2}+z}{y_{2}}\right)x_{2}^{-1}
\delta\left(\frac{x_{1}-x_{0}}{x_{2}}\right)\nno\\
&&=y_{1}^{-1}\delta\left(\frac{x_{1}+z}{y_{1}}\right)x_{2}^{-1}
\delta\left(\frac{y_{2}-z}{x_{2}}\right)x_{2}^{-1}
\delta\left(\frac{x_{1}-x_{0}}{x_{2}}\right)
\eea
and
\bea
\lefteqn{x_{1}^{-1}\delta\left(\frac{z-y_{1}}{-x_{1}}\right)x_{2}^{-1}
\delta\left(\frac{z-y_{2}}{-x_{2}}\right)y_{2}^{-1}
\delta\left(\frac{y_{1}-x_{0}}{y_{2}}\right)=}\nno\\
&&=z^{-1}\delta\left(\frac{-x_{1}+y_{1}}{z}\right)z^{-1}
\delta\left(\frac{-x_{2}+y_{2}}{z}\right)y_{2}^{-1}
\delta\left(\frac{y_{1}-x_{0}}{y_{2}}\right)\nno\\
&&=\left({\displaystyle \sum_{m, n\in {\Bbb Z}}}
\frac{(-x_{1}+y_{1})^{m}}{z^{m+1}}
\frac{(-x_{2}+y_{2})^{n}}{z^{n+1}}\right)
y_{2}^{-1}\delta\left(\frac{y_{1}-x_{0}}{y_{2}}\right)\nno\\
&&=\left({\displaystyle \sum_{m, n\in {\Bbb Z}}}(-x_{2}+y_{2})^{-1}
\left(\frac{-x_{1}+y_{1}}{-x_{2}+y_{2}}\right)^{m}\frac{(-x_{2}+y_{2})^{m+n+1}}
{z^{m+n+2}} \right)
y_{2}^{-1}\delta\left(\frac{y_{1}-x_{0}}{y_{2}}\right)\nno\\
&&=\left({\displaystyle \sum_{m, k\in {\Bbb Z}}}(-x_{2}+y_{2})^{-1}
\left(\frac{-x_{1}+y_{1}}{-x_{2}+y_{2}}\right)^{m}
z^{-1}\left(\frac{-x_{2}+y_{2}}{z}\right)^{k}\right)
y_{2}^{-1}\delta\left(\frac{y_{1}-x_{0}}{y_{2}}\right)\nno\\
&&=(-x_{2}+y_{2})^{-1}\delta\left(\frac{-x_{1}+y_{1}}{-x_{2}+y_{2}}\right)
z^{-1}\delta\left(\frac{-x_{2}+y_{2}}{z}\right)
y_{2}^{-1}\delta\left(\frac{y_{1}-x_{0}}
{y_{2}}\right)\nno\\
&&=(-x_{2})^{-1}\delta\left(\frac{x_{1}-(y_{1}-y_{2})}{x_{2}}\right)
z^{-1}\delta\left(\frac{-x_{2}+y_{2}}{z}\right)
y_{1}^{-1}\delta\left(\frac{y_{2}+x_{0}}
{y_{1}}\right)\nno\\
&&=x_{2}^{-1}\delta\left(\frac{x_{1}-x_{0}}{x_{2}}\right)x_{2}^{-1}
\delta\left(\frac{z-y_{2}}{-x_{2}}\right)y_{1}^{-1}\delta
\left(\frac{y_{2}+x_{0}}{y_{1}}\right),
\eea
(8.3) becomes
\bea
\lefteqn{([Y'_{Q(z)}(v_{1}, x_{1}), Y'_{Q(z)}(v_{2}, x_{2})]\lambda)(w_{(1)}
\otimes w_{(2)})=}\nno\\
&&=\mbox{\rm Res}_{x_{0}}\mbox{\rm Res}_{y_{2}}
\mbox{\rm Res}_{y_{1}}y_{1}^{-1}\delta\left(\frac{x_{1}+z}{y_{1}}\right)
x_{2}^{-1}\delta\left(\frac{y_{2}-z}{x_{2}}\right)x_{2}^{-1}
\delta\left(\frac{x_{1}-x_{0}}{x_{2}}\right)\cdot \nno\\
&&\hspace{4em}\cdot \lambda(
Y^{*}_{1}(Y(v_{1}, x_{0})v_{2}, y_{2})w_{(1)}\otimes w_{(2)})\nno\\
&&\quad -\mbox{\rm Res}_{x_{0}}
\mbox{\rm Res}_{y_{2}}\mbox{\rm Res}_{y_{1}}x_{2}^{-1}
\delta\left(\frac{x_{1}-x_{0}}{x_{2}}\right)x_{2}^{-1}
\delta\left(\frac{z-y_{2}}{-x_{2}}\right)y_{1}^{-1}
\delta\left(\frac{y_{2}+x_{0}}{y_{1}}\right)\cdot \nno\\
&&\hspace{4em}\cdot \lambda(w_{(1)}\otimes Y_{2}(Y(v_{1}, x_{0})v_{2},
y_{2})w_{(2)})\nno\\
&&=\mbox{\rm Res}_{x_{0}}x_{2}^{-1}\delta\left(\frac{x_{1}-x_{0}}
{x_{2}}\right)\cdot\nno\\
&&\hspace{2em}\cdot \biggl(\mbox{\rm Res}_{y_{2}}
x_{2}^{-1}\delta\left(\frac{y_{2}-z}
{x_{2}}\right)\lambda(
Y^{*}_{1}(Y(v_{1}, x_{0})v_{2}, y_{2})w_{(1)}\otimes w_{(2)})\nno\\
&&\hspace{4em}-\mbox{\rm Res}_{y_{2}}x_{2}^{-1}\delta\left(\frac{z-y_{2}}
{-x_{2}}\right)\lambda(w_{(1)}\otimes Y_{2}(Y(v_{1}, x_{0})v_{2}, y_{2})
w_{(2)})\biggr)
\nno\\
&&=\mbox{\rm Res}_{x_{0}}x_{2}^{-1}\delta\left(\frac{x_{1}-x_{0}}{x_{2}}\right)
(Y'_{Q(z)}(Y(v_{1}, x_{0})v_{2}, x_{2})\lambda)(w_{(1)}\otimes w_{(2)}).
\eea
Since $\lambda$, $w_{(1)}$ and $w_{(2)}$ are arbitrary,
this  equality gives the commutator formula for $Y'_{Q(z)}$,  completing
the proof of Proposition 5.2.

\begin{rema}
{\rm The proof of Proposition 5.2 suggests the following:  Using
the definitions (5.1) and (5.2) as motivation, let us define a (linear) action
$\sigma_{Q(z)}$ of $V\otimes \iota_{+}{\Bbb C}[t, t, (z+t)^{-1}]$ on the
vector space $W_{1}\otimes W_{2}$ (as opposed to $(W_{1}\otimes W_{2})^{*}$)
as follows:
\begin{equation}
\sigma_{Q(z)}(\xi)(w_{(1)}\otimes w_{(2)})
=\tau_{W_{1}}(T_{-z}^{*}\xi)w_{(1)}\otimes w_{(2)}
-w_{(1)}\otimes \tau_{W_{2}}(T_{-z}^{+}\xi)w_{(2)}
\end{equation}
for $\xi\in V\otimes \iota_{+} {\Bbb C}[t, t^{-1}, (z+t)^{-1}]$, $w_{(1)}\in
W_{1}$, $w_{(2)}\in W_{2}$, or equivalently,
\begin{eqnarray}
\lefteqn{\sigma_{Q(z)}
\left(z^{-1}\delta\left(\frac{x_{1}-x_{0}}{z}\right)
Y_{t}(v, x_{0})\right)(w_{(1)}\otimes w_{(2)})}\nonumber\\
&&=x^{-1}_{0}\delta\left(\frac{x_{1}-z}{x_{0}}\right)
Y_{1}^{*}(v, x_{1})w_{(1)}\otimes w_{(2)}\nonumber\\
&&\hspace{2em}-x_{0}^{-1}\delta\left(\frac{z-x_{1}}{-x_{0}}\right)
w_{(1)}\otimes Y_{2}(v, x_{1})w_{(2)}.
\end{eqnarray}
That is, the operators $\sigma_{Q(z)}(\xi)$ and $\tau_{Q(z)}(\xi)$ are
mutually adjoint:
\begin{equation}
(\tau_{Q(z)}(\xi)\lambda)(w_{(1)}\otimes w_{(2)})
=\lambda(\sigma_{Q(z)}(\xi)(w_{(1)}\otimes w_{(2)})).
\end{equation}
While this action on $W_{1}\otimes W_{2}$ is not very useful, it has the
following three properties:
\begin{equation}
\sigma_{Q(z)}(Y_{t}({\bf 1}, x))=1,
\end{equation}
\begin{equation}
\frac{d}{dx}\sigma_{Q(z)}(Y_{t}(v, x))=\sigma_{Q(z)}(Y_{t}(L(-1)v, x))
\end{equation}
for $v\in V$,
\begin{eqnarray}
\lefteqn{[\sigma_{Q(z)}(Y_{t}(v_{2}, x_{2})), \sigma_{Q(z)}(Y_{t}(v_{1},
x_{1}))]}\nno\\
&&=\res_{x_{0}}x_{2}^{-1}\delta\left(\frac{x_{1}-x_{0}}{x_{2}}\right)
\sigma_{Q(z)}(Y_{t}(Y(v_{1}, x_{0})v_{2}, x_{2}))
\end{eqnarray}
for $v_{1}, v_{2}\in V$ (the {\it opposite commutator formula}). These
follow either {}from the assertions of Propositions 5.1 and 5.2 or,
better, {}from the fact that it was actually (8.10)--(8.12) that the
proofs of these propositions were proving.}
\end{rema}

\renewcommand{\theequation}{\thesection.\arabic{equation}}
\renewcommand{\therema}{\thesection.\arabic{rema}}
\setcounter{equation}{0}
\setcounter{rema}{0}

\section{Some useful formulas}

In this section, we shall prove some formulas which will be useful
in the next  section.

Let $\lambda\in (W_{1}\otimes W_{2})^{*}$, $w_{(1)}\in W_{1}$ and
$w_{(2)}\in W_{2}$.
{}From the definitions of $L'_{Q(z)}(0)$ and of $\tau_{Q(z)}$
(see Section 5), we have
\begin{eqnarray}
\lefteqn{(L'_{Q(z)}(0)\lambda)(w_{(1)} \otimes w_{(2)})=}\nno\\
&&=(\tau_{Q(z)}(\omega\otimes
t) \lambda)(w_{(1)} \otimes w_{(2)})\nno\\
&&= \lambda((L(0) -
zL(1))w_{(1)} \otimes w_{(2)})- \lambda(w_{(1)} \otimes (L(0) -
zL(-1))w_{(2)}),\;\;\;\;\;
\end{eqnarray}
where we have used the same notations $L(0), L(-1), L(1)$ to denote the
actions of
the corresponding elements of the Virasoro algebra on both $W_1$
and $W_2$.
For conveience we write $L(-1)=L'_{Q(z)}(-1)$, $L(0) = L'_{Q(z)}(0)$ and
$L(1)=L'_{Q(z)}(1)$ in the rest of this paper.
There
will be no confusion since the three $L(0)$'s act on different
spaces.
\begin{lemma}
For $\lambda\in (W_{1}\otimes W_{2})^{*}$, $w_{(1)}\in W_{1}$
and $w_{(2)}\in W_{2}$,
we have
\begin{eqnarray}
\lefteqn{\biggl(\biggl(1-\frac{y_1}{z}\biggr)^{L(0)} \lambda\biggr)(w_{(1)}
\otimes w_{(2)})=}\nno\\
&&= \lambda\biggl(\biggl(1-\frac{y_1}{z}\biggr)^{L(0)-zL(1)}w_{(1)}
\otimes \biggl(1-\frac{y_1}{z}\biggr)^{-(L(0)-z L(-
1))}w_{(2)}\biggr).
\end{eqnarray}
\end{lemma}
 \pf
{}From (9.1) we have
\begin{eqnarray}
\lefteqn{\biggl(\biggl(1-\frac{y_1}{z}\biggr)^{L(0)} \lambda)(w_{(1)}
\otimes w_{(2)})=}\nno\\
&&= (e^{L(0)\log(1-\frac{y_1}{z})} \lambda)(w_{(1)}
\otimes w_{(2)})\nno\\
&&= \lambda(e^{L(0)-z L(1))\log(1-\frac{y_1}{z})}w_{(1)}
\otimes e^{-(L(0)-z L(-1))\log(1-\frac{y_1}{z})}w_{(2)})\nno\\
&&= \lambda\biggl(\biggl(1-\frac{y_1}{z}\biggr)^{L(0)-zL(1)}w_{(1)}
\otimes \biggr(1-\frac{y_1}{z}\biggr)^{-(L(0)-z L(-
1))}w_{(2)}\biggr). \hspace{1.5em}\Box
\end{eqnarray}

\begin{lemma}
For $v\in V$,
\begin{equation}
Y'_{Q(z)}(v,x) = \biggl(1-
\frac{y_1}{z}\biggr)^{L(0)}Y'_{Q(z)}
\biggl(\biggl(1-\frac{y_1}{z}\biggr)^{-L(0)}
v,\frac{x}{1-y_1/z}\biggr)\biggr(1-\frac{y_1}{z}\biggr)^{-
L(0)}.
\end{equation}
This formula also holds for the vertex operators associated with any
$V$-module.
\end{lemma}
\pf
The identity (9.4) will follow {}from
\begin{equation}
e^{yL(0)}Y'_{Q(z)}(v, x)e^{-yL(0)}=Y'_{Q(z)}(e^{yL(0)}
v, e^{y}x).
\end{equation}
This is analogous to formula (5.2.37) of \cite{FHL}, but it is different
because here we have $e^y$, not $y$; also, the proof of
(5.2.37) in \cite{FHL} uses the grading of the module, which we do not have
available here.  To prove (9.5) assume without loss of generality that
 $\mbox{wt}\ v=h\in {\Bbb Z}$, and use
the $L(-1)$-derivative property (5.6) and the commutator formula
(Proposition 5.2) to get
\begin{equation}
[L(0), Y'_{Q(z)}(v, x)]=\left(x\frac{d}{dx}+h\right)Y'_{Q(z)}(v, x),
\end{equation}
 which is the case $n=0$ of  (2.6.2) of \cite{FHL}.  Exactly as
in Section 2.6 of \cite{FHL}, we obtain the desired result.\epf

\begin{lemma}
Let $L(-1)$, $L(0)$ be two operators satisfying the commutator relation
\begin{equation}
[L(0), L(-1)]=L(-1).
\end{equation}
 Then
\begin{equation}
\biggr(1-\frac{y_1}{z}\biggr)^{L(0)-zL(-1)}=e^{y_1L(-1)}
\biggr(1-\frac{y_1}{z}\biggr)^{L(0)}.
\end{equation}
\end{lemma}
\pf
 We first prove that the derivative of
$$(1-x)^{L(0)-zL(-1)}
(1-x)^{-L(0)}
e^{-zxL(-1)}$$ is $0$.  Write $A=(1-x)^{L(0)-zL(-1)}$,
$B=(1-x)^{-L(0)}$,
$C=e^{-zxL(-1)}$.  Then
 \begin{eqnarray}
\frac{d}{dx}(ABC)
&=&-A(1-x)^{-1}(L(0)-zL(-1)))BC\nno\\
&&+A(1-x)^{-1}L(0)BC\nno\\
&&-zABL(-1)C.
\end{eqnarray}
Using (5.2.12) of \cite{FHL} we have
\begin{eqnarray}
BL(-1)&=&(1-x)^{-L(0)}L(-1)\nno\\
&=&e^{(-\log(1-x))L(0)}L(-1)\nno\\
&=&L(-1)e^{(-\log(1-x))L(0)}e^{-\log(1-x)}\nno\\
&=&L(-1)(1-x)^{-L(0)}
(1-x)^{-1}\nno\\
&=&(1-x)^{-1}L(-1)B.
\end{eqnarray}
Substituting (9.10) into (9.9), we obtain
\begin{eqnarray*}
\frac{d}{dx}(ABC)&=&-A(1-x)^{-1}L(0)BC
+zA(1-x)^{-1}L(-1)BC
\nno\\
&&+A(1-x)^{-1}L(0)BC
-zA(1-x)^{-1}L(-1)BC\nno\\
&=&0.
\end{eqnarray*}
Thus $ABC$ is constant, and since
$ABC\lbar_{x=0}=1$, we have $ABC=1$,
which is equivalent to (9.8).\epf

\renewcommand{\theequation}{\thesection.\arabic{equation}}
\renewcommand{\therema}{\thesection.\arabic{rema}}
\setcounter{equation}{0}
\setcounter{rema}{0}

\section{Jacobi identity for vertex operators  on a certain subspace of
$(W_{1}\otimes W_{2})^{*}$}

We prove Theorem 6.1  in this section. The reader should again observe the
justifiability of each formal step in the argument; sometimes this is quite
subtle.

Let $\lambda$ be an element of  $(W_{1}\otimes W_{2})^{*}$
satisfying the compatibility condition, that is,
(a) the lower truncation condition: for all $v\in V$,
\begin{equation}
\tau_{Q(z)}(v \otimes t^n)\lambda = 0 \ \ \mbox{for}\ n \
\mbox{sufficiently\ large},
\end{equation}
or equivalently, for all $v \in V$,
\begin{eqnarray}
Y'_{Q(z)}(v,x)\lambda& =& \tau_{Q(z)}(Y_t(v,x))\lambda\nno\\
&= &\sum_{n < N} \tau_{Q(z)}(v \otimes t^n)\lambda x^{-n-1}\ \ \mbox{for\
some}\ \ N \in {\Bbb Z},
\end{eqnarray}
 and  (b) the formula
\begin{eqnarray}
\lefteqn{
\tau_{Q(z)}\left(x^{-1}_1\delta\left(\frac{z+x_0}{x_1}\right)Y_t(v,x_0)
\right)\lambda=}\nno\\
&&=x^{-1}_1
\delta\left(\frac{z+x_0}{x_1}\right)Y'_{Q(z)}(v,x_0)\lambda\;\;\;
\mbox{\rm for all}\ v\in V.
\end{eqnarray}
By (5.2) and (5.4), (10.3) is equivalent to
\begin{eqnarray}
\lefteqn{x^{-1}_0 \delta\left(\frac{x_1-
z}{x_0}\right)\lambda(Y^*_1(v,x_1)w_{(1)} \otimes w_{(2)})}\nno\\
&&\hspace{2em}- x^{-1}_0 \delta\left(\frac{z-x_1}{-x_0}\right)\lambda(w_{(1)}
\otimes
Y_2(v,x_1)w_{(2)})\nno\\
&&= x^{-1}_1
\delta\left(\frac{z+x_0}{x_1}\right)\biggl(\res_{y_{1}}x^{-1}_0
\delta\left(\frac{y_1-
z}{x_0}\right)\lambda(Y^*_1(v,y_1)w_{(1)} \otimes
w_{(2)})\nno\\
&&\hspace{2em}- \res_{y_1}x^{-1}_0 \delta\left(\frac{z-y_1}{-
x_0}\right)\lambda(w_{(1)} \otimes Y_2(v,y_1)w_{(2)})\biggr)
\end{eqnarray}
for all $v\in V$, $w_{(1)}\in W_{1}$ and $w_{(2)}\in W_{2}$.
Note that on the right-hand side the distributive law is not valid since the
two individual products are not defined. One feature of the argument which
follows is that we must rewrite expressions to allow the application of
distributivity.

By (5.4), we have
\begin{eqnarray}
\lefteqn{\left(x^{-
1}_0\delta\left(\frac{x_1-x_2}{x_0}\right)
Y'_{Q(z)}(v_1,x_1)Y'_{Q(z)}(v_2,x_2)
\lambda\right)(w_{(1)} \otimes w_{(2)})}\nno\\
&&= x^{-1}_0 \delta\left(\frac{x_1-
x_2}{x_0}\right)(Y'_{Q(z)}(v_1,x_1)Y'_{Q(z)}(v_2,x_2) \lambda)(w_{(1)} \otimes
w_{(2)})\nno\\
&&= x^{-1}_0 \delta\left(\frac{x_1-x_2}{x_0}\right)\biggl(\res_{y_1}x^{-
1}_1 \delta\left(\frac{y_1-z}{x_1}\right)\cdot\nno\\
&&\hspace{6em}\cdot (Y'_{Q(z)}(v_2,x_2)
\lambda)(Y^*_1(v_1,y_1)w_{(1)} \otimes w_{(2)})\nno\\
 &&\quad -\res_{y_1} x^{-
1}_1 \delta \left(\frac{z - y_1}{-x_1}\right)(Y'_{Q(z)}(v_2,x_2)
\lambda)(w_{(1)} \otimes Y_2(v_1,y_1)w_{(2)})\biggr)\nno\\
&&= x^{-1}_0
\delta\left(\frac{x_1-x_2}{x_0}\right)\biggl(\res_{y_1}x^{-1}_1
\delta\left(\frac{y_1-
z}{x_1}\right)\res_{y_2}x^{-1}_2 \delta \left(\frac{y_2-
z}{x_2}\right)\cdot\nno\\
&&\hspace{6em}\cdot \lambda(Y^*_1(v_2,y_2)Y^*_1(v_1,y_1)w_{(1)} \otimes
w_{(2)})\nno\\
&&\quad -\res_{y_1}x^{-1}_1 \delta\left(\frac{y_1-
z}{x_1}\right)\res_{y_2}x^{-1}_2 \delta\left(\frac{z-y_2}{-
x_2}\right)\cdot \nno\\
&&\hspace{6em}\cdot \lambda(Y^*_1(v_1,y_1)w_{(1)} \otimes Y_2(v_2,y_2)
w_{(2)})\nno\\
&&\quad -
\res_{y_1}x^{-1}_1 \delta\left(\frac{z-y_1}{-x_1}\right)(Y'_{Q(z)}(v_2,x_2)
\lambda)(w_{(1)} \otimes Y_2(v_1,y_1)w_{(2)})\biggr).\;\;\;\;\;\;
\end{eqnarray}
{}From the properties of the formal $\delta$-function,
we see that the right-hand side of (10.5)
is equal to
\begin{eqnarray}
\lefteqn{x^{-1}_0
\delta\left(\frac{x_1-x_2}{x_0}\right)\biggl(\res_{y_1}y^{-1}_1
\delta\left(\frac{x_1+z}{y_1}\right)\res_{y_2}y^{-1}_2
\delta\left(\frac{x_2+z}{y_2}\right)}\cdot \nno\\
&&\hspace{6em}\cdot \lambda(Y^*_1(v_2,x_2+z)Y^*_1(v_1,x_1+z)w_{(
1)} \otimes w_{(2)})\nno\\
&&\quad -\res_{y_1}y^{-1}_1
\delta\left(\frac{x_1+z}{y_1}\right)\res_{y_2}x^{-1}_2
\delta\left(\frac{z-y_2}{-
x_2}\right)\cdot \nno\\
&&\hspace{6em}\cdot \lambda(Y^*_1(v_1,x_1 +z)w_{(1)} \otimes
Y_2(v_2,y_2)w_{(2)})\nno\\
&&\quad - \res_{y_1}x^{-1}_1 \delta\left(\frac{z-y_1}{-
x_1}\right)(Y'_{Q(z)}(v_2,x_2) \lambda)(w_{(1)} \otimes
Y_2(v_1,y_1)w_{(2)})\biggr)\nno\\
&&= x^{-1}_0 \delta \left(\frac{x_1-
x_2}{x_0}\right)\biggl(\lambda(Y^*_1(v_2,x_2+z)
Y^*_1(v_1,x_1+z)w_{(1)}
\otimes w_{(2)})\nno\\
&&\quad - \res_{y_2}x^{-1}_2 \delta\left(\frac{z-y_2}{-
x_2}\right)\lambda(Y^*_1(v_1,x_1+z)w_{(1)} \otimes
Y_2(v_2,y_2)w_{(2)})\nno\\
&&\quad - \res_{y_1}x^{-1}_1 \delta\left(\frac{z-y_1}{-
x_1}\right)(Y'_{Q(z)}(v_2,x_2) \lambda)(w_{(1)} \otimes
Y_2(v_1,y_1)w_{(2)})\biggr).\;\;\;\;\;\;\;\;\;
\end{eqnarray}
{}From the $L(-1)$-derivative property for $Y_{1}$ and $Y_{1}^{*}$ and
the commutator formulas for $L(-1)$, $Y_{1}(\cdot, x)$ and for $L(1)$,
$Y^{*}_{1}(\cdot, x)$, we obtain
\begin{eqnarray}
Y_{1}(v, x+z)&=&Y_{1}(e^{zL(-1)}v, x)\nno\\
&=&e^{zL(-1)}Y_{1}(v, x)e^{-zL(-1)}\nno\\
&=&\sum_{n\ge
0}\frac{z^{n}}{n!}\frac{d^{n}}{dx^{n}}Y_{1}(v, x),
\end{eqnarray}
and
\begin{eqnarray}
Y^{*}_{1}(v, x+z)&=&Y^{*}_{1}(e^{zL(-1)}v, x)\nno\\
&=&e^{-zL(1)}Y^{*}_{1}(v, x)e^{zL(1)}\nno\\
&=&\sum_{n\ge
0}\frac{z^{n}}{n!}\frac{d^{n}}{dx^{n}}Y^{*}_{1}(v, x).
\end{eqnarray}
(Note that all these expressions are in fact defined.)
Using (10.8), we see that the right-hand side of (10.6) can be written as
\begin{eqnarray}
\lefteqn{x^{-1}_0 \delta \left(\frac{x_1-
x_2}{x_0}\right)\biggl(\lambda(Y^*_1(e^{zL(-1)}v_2,x_2)
Y^*_1(e^{zL(-1)}v_1,x_1)w_{(1)}
\otimes w_{(2)})}\nno\\
&&\quad - \res_{y_2}x^{-1}_2 \delta\left(\frac{z-y_2}{-
x_2}\right)\lambda(Y^*_1(e^{zL(-1)}v_1,x_1)w_{(1)} \otimes
Y_2(v_2,y_2)w_{(2)})\nno\\
&&\quad - \res_{y_1}x^{-1}_1 \delta\left(\frac{z-y_1}{-
x_1}\right)(Y'_{Q(z)}(v_2,x_2) \lambda)(w_{(1)} \otimes
Y_2(v_1,y_1)w_{(2)})\biggr)\nno\\
&&=x^{-1}_0 \delta \left(\frac{x_1-
x_2}{x_0}\right)\biggl(\lambda(e^{-zL(1)}Y^*_1(v_2,x_2)Y^*_1(v_1,x_1)e^{zL(1)}
w_{(1)}
\otimes w_{(2)})\nno\\
&&\quad - \res_{y_2}x^{-1}_2 \delta\left(\frac{z-y_2}{-
x_2}\right)\lambda(e^{-zL(1)}Y^*_1(v_1,x_1)e^{zL(1)}w_{(1)} \otimes
Y_2(v_2,y_2)w_{(2)})\nno\\
&&\quad - \res_{y_1}x^{-1}_1 \delta\left(\frac{z-y_1}{-
x_1}\right)(Y'_{Q(z)}(v_2,x_2) \lambda)(w_{(1)} \otimes
Y_2(v_1,y_1)w_{(2)})\biggr).
\end{eqnarray}
Note that it is easier to verify the well-definedness of the terms on
the right-hand side of (10.6) than that of the terms in (10.9), though
(10.9) is sometimes easier to use if it is known that every term is
well defined.  Below we shall write expressions like those on the right-hand
side of (10.6) in whichever way suits our needs.
The distributive law applies to the right-hand side of (10.6) (or (10.9)) since
all three of the following expressions are defined: $$x^{-1}_0
\delta\left(\frac{x_1-x_2}{x_0}\right)\lambda(Y^*_1(v_2,x_2+z)
Y^*_1(v_1,x_1+z)w_{(1)} \otimes w_{(2)}),$$ $$x^{-1}_0
\delta\left(\frac{x_1-x_2}{x_0}\right)\res_{y_2}x^{-1}_2
\delta\left(\frac{z-y_2}{-x_2}\right)\lambda(Y^*_1(v_1,x_1+z)w_{(1)} \otimes
Y_2(v_2,y_2)w_{(2)}),$$
$$x^{-1}_0
\delta\left(\frac{x_1-x_2}{x_0}\right)\res_{y_1}x^{-1}_1 \delta\left(\frac{z-
y_1}{-x_1}\right)(Y'_{Q(z)}(v_2,x_2) \lambda)(w_{(1)} \otimes
Y_2(v_1,y_1)w_{(2)}).$$

Now we examine the last expression in (10.9).
Rewriting the formal $\delta$-functions
$x_{0}^{-1}\delta\left(\frac{x_{1}-x_{2}}{x_{0}}\right)$ and
$x_{1}^{-1}\delta\left(\frac{z-y_1}{-x_1}\right)$, and using (9.4) and the
fundamental property of the $\delta$-function, we have:
\begin{eqnarray}
\lefteqn{x^{-1}_0\delta\left(\frac{x_1-x_2}{x_0}\right)\res_{y_{1}}x^{-1}_1
\delta\left(\frac{z-y_1}{-x_1}\right)\cdot}\nno\\
&&\hspace{4em}\cdot (Y'_{Q(z)}(v_2, x_2)\lambda)(w_{(1)}\otimes
 Y_2(v_1,
y_1)w_{(2)})\nno\\
&&=\biggl(\frac{x_1}{z}\biggr)^{-1}\biggl(\frac{x_0}{x_1/z}\biggr)^{-
1}\delta\left(\frac{z+(\frac{x_2}{-x_1/z})}{\frac{x_0}{x_1/z}}\right)
\res_{y_1}x_1^{-1}\delta\left(\frac{1-y_1/z}{-x_1/z}\right)\cdot \nno\\
&&\hspace{4em}\cdot\biggl(\biggl(1-\frac{y_1}{z}\biggr)^{L(0)}Y'_{Q(z)}
\biggl(\biggl(1-\frac{y_1}{z}\biggr)^{-
L(0)}v_2, \frac{x_2}{1-y_{1}/z}\biggr)\cdot \nno\\
&&\hspace{5em}\cdot \biggl(1-\frac{y_1}{z}\biggr)^{-L(0)}
\lambda\biggr)(w_{(1)}\otimes Y_2(v_1, y_1)w_{(2)})\nno\\
&&=\biggl(\frac{x_1}{z}\biggr)^{-1}\biggl(\frac{x_0}{x_1/z}\biggr)^{-
1}\delta\left(\frac{z+(\frac{x_2}{-x_1/z})}{\frac{x_0}{x_1/z}}\right)
\res_{y_1}x_1^{-1}\delta\left(\frac{1-y_1/z}{-x_1/z}\right)\cdot \nno\\
&&\hspace{4em}\cdot\biggl(\biggl(1-\frac{y_1}{z}\biggr)^{L(0)}Y'_{Q(z)}
\biggl(\biggl(1-\frac{y_1}{z}\biggr)^{-
L(0)}v_2, \frac{x_2}{-x_{1}/z}\biggr)\cdot \nno\\
&&\hspace{5em}\cdot \biggl(1-\frac{y_1}{z}\biggr)^{-L(0)}
\lambda\biggr)(w_{(1)}\otimes Y_2(v_1, y_1)w_{(2)}).
\end{eqnarray}
By (9.2) and (10.3),  the right-hand
side of (10.10) is equal to
\begin{eqnarray}
&&\biggl(\frac{x_1}{z}\biggr)^{-1}\biggl(\frac{x_0}{x_1/z}\biggr)^{-
1}\delta\left(\frac{z+(\frac{x_2}{-x_1/z})}{\frac{x_0}{x_1/z}}\right)
\res_{y_1}x_1^{-1}\delta\left(\frac{1-y_1/z}{-x_1/z}\right)\cdot \nno\\
&&\hspace{2em}\cdot
\biggl(Y'_{Q(z)}\biggl(\biggl(1-\frac{y_1}{z}\biggr)^{-L(0)}
v_2, \frac{x_2}{-x_1/z}\biggr)\biggl(1-
\frac{y_1}{z}\biggr)^{-L(0)} \lambda\biggr)\cdot \nno\\
&&\hspace{3em}\cdot \biggl(\biggl(1-\frac{y_1}{z}\biggr)^{L(0)-
zL(1)}w_{(1)}\otimes\nno\\
&&\hspace{6em}\otimes\biggl(1-\frac{y_1}{z}\biggr)^{-
L(0)-zL(-1)}Y_2(v_1, y_1)w_{(2)}\biggr)\nno\\
&&=\biggl(\frac{x_1}{z}\biggr)^{-1}
\res_{y_1}x_1^{-1}\delta\left(\frac{1-y_1/z}{-x_1/z}\right)\cdot \nno\\
&&\hspace{2em} \cdot\Biggl(\tau_{Q(z)}\Biggl(\biggl(\frac{x_0}{x_1/z}
\biggr)^{-
1}\delta\left(\frac{z+(\frac{x_2}{-x_1/z})}{\frac{x_0}{x_1/z}}\right)
\cdot \nno\\
&&\hspace{3em}\cdot Y_{t}\biggl(\biggl(1-\frac{y_1}{z}\biggr)^{-L(0)}v_2,
\frac{x_2}{-x_1/z}\biggr)\Biggr)\biggl(1-
\frac{y_1}{z}\biggr)^{-L(0)} \lambda\Biggr)\cdot \nno\\
&&\hspace{4em}\cdot \biggl(\biggl(1-\frac{y_1}{z}\biggr)^{L(0)-
zL(1)}w_{(1)}\otimes\nno\\
&&\hspace{6em}\otimes\biggl(1-\frac{y_1}{z}\biggr)^{-
L(0)-zL(-1)}Y_2(v_1, y_1)w_{(2)}\biggr).
\end{eqnarray}
By (5.2), the right-hand side of (10.11) becomes
\begin{eqnarray}
&&\biggl(\frac{x_1}{z}\biggr)^{-1}
\res_{y_1}x_1^{-1}\delta\left(\frac{1-y_1/z}{-x_1/z}\right)
\biggl(\biggl(\frac{x_2}{-x_1/z}\biggr)^{-
1}\delta\left(\frac{\frac{x_0}{x_1/z}-z}{\frac{x_2}{-x_1/z}}\right)\cdot \nno\\
&&\hspace{2em}\cdot \biggl(\biggl(1-\frac{y_1}{z}\biggr)^{-L(0)}\lambda\biggr)
\biggl(Y_1^*\biggl(\biggl(1-\frac{y_1}{z}\biggr)^{-L(0)}v_2,\frac{x_0}
{x_{1}/z}\biggr)\cdot \nno\\
&&\hspace{3em}\cdot \biggl(1-\frac{y_1}{z}\biggr)^{L(0)-zL(1)}
w_{(1)}\otimes \biggl(1-\frac{y_1}{z}\biggr)^{-(L(0)-zL(1))}Y_2(v_1, y_1)
w_{(2)}\biggr)\nno\\
&&\quad -\biggl(\frac{x_2}{-x_1/z}\biggr)^{-
1}\delta\left(\frac{z-\frac{x_0}{x_1/z}}{-\frac{x_2}{-x_1/z}}\right)
\biggl(\biggl(1-\frac{y_1}{z}\biggr)^{-L(0)}\lambda\biggr)
\biggl(\biggl(1-\frac{y_1}{z}\biggr)^{L(0)-zL(1)} \cdot\nno\\
&&\hspace{2em}\cdot w_{(1)} \otimes
 Y_2\biggl(\biggl(1-\frac{y_1}{z}\biggr)^{-L(0)}v_2,
\frac{x_0}{x_1/z}\biggr)\cdot\nno\\
&&\hspace{3em}\cdot
\biggl(1-\frac{y_1}{z}\biggr)^{(L(0)-zL(-1))}Y_2(v_1,
y_1)w_{(2)}\biggr)\biggr).\;\;\;\;\;\;\;
\end{eqnarray}
Using (9.2) again but with $1-\frac{y_{1}}{z}$ replaced by
$(1-\frac{y_{1}}{z})^{-1}$, rewriting formal
$\delta$-functions and then using the distributive law,
we see that (10.12) is equal to
\begin{eqnarray}
&&\res_{y_1}x_1^{-1}\delta\left(\frac{1-y_1/z}{-x_1/z}\right)
\biggl(-x_2^{-1}\delta\left(\frac{x_0-x_1}{-x_2}\right)\cdot \nno\\
&&\hspace{4em}\cdot \lambda\biggl(\biggl(1-\frac{y_1}{z}\biggr)^{-(L(0)-zL(1))}
Y_1^*\biggl(\biggl(1-\frac{y_1}{z}\biggr)^{-L(0)}v_2,\frac{x_0}{-(1-
y_1/z)}\biggr)\cdot \nno\\
&&\hspace{5em}\cdot \biggl(1-\frac{y_1}{z}\biggr)^{L(0)-zL(1)}
w_{(1)}\otimes Y_2(v_1, y_1)w_{(2)}\biggr)\nno\\
&&\quad +x_2^{-1}\delta\left(\frac{x_1-x_0}{x_2}\right)
\lambda(w_{(1)}\otimes\biggl(1-\frac{y_1}{z}\biggr)^{L(0)-zL(-1)}\cdot \nno\\
&&\hspace{4em}\cdot Y_2\biggl(\biggl(1-\frac{y_1}{z}\biggr)^{-L(0)}v_2,
\frac{x_0}{-(1-
y_1/z)}\biggr)\cdot \nno\\
&&\hspace{5em}\cdot \biggl(1-\frac{y_1}{z}\biggr)^{-
(L(0)-zL(-1))}Y_2(v_1, y_1)w_{(2)})\biggr)\nno\\
&&=-\res_{y_1}x_1^{-1}\delta\left(\frac{1-y_1/z}{-x_1/z}\right)
x_2^{-1}\delta\left(\frac{x_0-x_1}{-x_2}\right)\cdot \nno\\
&&\hspace{4em}\cdot \lambda\biggl(\biggl(1-\frac{y_1}{z}\biggr)^{-(L(0)-zL(1))}
Y_1^*\biggl(\biggl(1-\frac{y_1}{z}\biggr)^{-L(0)}v_2,\frac{x_0}{-(1-
y_1/z)}\biggr)\cdot \nno\\
&&\hspace{5em}\cdot \biggl(1-\frac{y_1}{z}\biggr)^{L(0)-zL(1)}
w_{(1)}\otimes Y_2(v_1, y_1)w_{(2)}\biggr)\nno\\
&&\quad +\res_{y_1}x_1^{-1}\delta\left(\frac{1-y_1/z}{-x_1/z}\right)
x_2^{-1}\delta\left(\frac{x_1-x_0}{x_2}\right)
\lambda(w_{(1)}\otimes\nno\\
&&\hspace{4em}\otimes\biggl(1-\frac{y_1}{z}\biggr)^{L(0)-zL(-1)}
Y_2\biggl(\biggl(1-\frac{y_1}{z}\biggr)^{-L(0)}v_2,
\frac{x_0}{-(1-
y_1/z)}\biggr)\cdot \nno\\
&&\hspace{5em}\cdot \biggl(1-\frac{y_1}{z}\biggr)^{-
(L(0)-zL(-1))}Y_2(v_1, y_1)w_{(2)}).
\end{eqnarray}

But by (9.8) and (9.4),
\begin{eqnarray}
\lefteqn{\biggl(1-\frac{y_1}{z}\biggr)^{L(0)-zL(-1)} Y_2
\biggl(\biggl(1-\frac{y_1}{z}\biggr)^{-L(0)}v_2, \frac{x_0}{-(1-
y_1/z)}\biggr)\biggl(1-\frac{y_1}{z}\biggr)^{-(L(0)-zL(-
1))}}\nno\\
&&=e^{y_{1}L(-1)}\biggl(1-\frac{y_1}{z}\biggr)^{L(0)}Y_2
\biggl(\biggl(1-\frac{y_1}{z}\biggr)^{-L(0)}v_2, \frac{x_0}{-(1-
y_1/z)}\biggr)\cdot\hspace{4em}\nno\\
&&\hspace{10em}\cdot\biggl(1-\frac{y_1}{z}\biggr)^{-L(0)}e^{-y_{1}L(-1)}\nno\\
&&=e^{y_{1}L(-1)}Y_2(v_2, -x_0) e^{-y_{1}L(-1)}\nno\\
&&=Y_2(v_2, -x_0+ y_1)\nno\\
&&=Y_2(v_2, -x_0-(z-y_1)+z)\nno\\
&&=Y_2(e^{zL(-1)}v_2, -x_0-
(z-y_1)).\hspace{17em}
\end{eqnarray}
We similarly have (using (9.8) and (9.4) for $Y'_1(v_2, x)$ and
 then using (3.22))
\begin{eqnarray}
\lefteqn{\biggl(1-\frac{y_1}{z}\biggr)^{-(L(0)-zL(1))}
Y_1^*\biggl(\biggl(1-\frac{y_1}{z}\biggr)^{-L(0)}v_2,\frac{x_0}{-(1-
y_1/z)}\biggr)\biggl(1-\frac{y_1}{z}\biggr)^{L(0)-zL(1)}}\nno\\
&&=e^{-y_{1}L(1)}Y_{1}^{*}(v_{2}, -x_{0})e^{y_{1}L(1)}\nno\\
&&=Y_{1}^{*}(v_{2}, -x_{0}+y_{1})\nno\\
&&=Y_{1}^{*}(v_{2}, -x_{0}-(z-y_{1})+z)\nno\\
&& =Y_1^*(e^{zL(-1)}v_2, -x_0-
(z-y_1)).\hspace{17em}
\end{eqnarray}
Substituting (10.14) and (10.15) into the right-hand side of (10.13) and then
combining  with  (10.10)--(10.13),
we obtain
\begin{eqnarray}
\lefteqn{x^{-1}_0\delta\left(\frac{x_1-x_2}{x_0}\right)\res_{y_1}
x^{-1}_1\delta\left(\frac{z-y_1}{-x_1}\right)(Y'_{Q(z)}(v_2, x_2) \lambda)
(w_{(1)}\otimes Y_2(v_1, y_1)w_{(2)})}\nno\\
&&=-\res_{y_1} x^{-1}_1\delta\left(\frac{z-y_1}{-x_1}\right)
x^{-1}_2\delta\left(\frac{x_0-x_1}{-x_2}\right)\cdot \nno\\
&&\hspace{4em}\cdot \lambda(Y_1^*\biggl(v_2,
-x_0-(z-y_1)+z\biggr)
w_{(1)}\otimes Y_2(v_1, y_1)w_{(2)})\nno\\
&&\quad +\res_{y_1} x^{-1}_1\delta\left(\frac{z-y_1}{-x_1}\right)
x^{-1}_2\delta\left(\frac{x_1-x_0}{x_2}\right)\cdot \nno\\
&&\hspace{4em}\cdot \lambda(w_{(1)}\otimes Y_2(v_2, -x_0+y_1)
Y_2(v_1, y_1)w_{(2)}).\hspace{8em}
\end{eqnarray}
(We choose the form of the expression {}from (10.15) in anticipation of the
next step.)

By the  properties of the formal $\delta$-function, the
right-hand side of (10.16) is equal to
\begin{eqnarray}
&&-x^{-1}_2\delta\left(\frac{-x_0+x_1}{x_2}\right)\res_{y_1}
x^{-1}_1\delta\left(\frac{z-y_1}{-x_1}\right)\cdot \nno\\
&&\hspace{4em}\cdot \lambda(Y_1^*(v_2, -x_0+x_1+z)w_{(1)}\otimes Y_2(v_1,
y_1)w_{(2)})\nno\\
&&\quad +x^{-1}_2\delta\left(\frac{x_1-x_0}{x_2}\right)\res_{y_1}
x^{-1}_1\delta\left(\frac{z-y_1}{-x_1}\right)\cdot \nno\\
&&\hspace{4em}\cdot \lambda(w_{(1)}\otimes Y_2(v_2, -x_0+y_1)
Y_2(v_1, y_1)w_{(2)})\nno\\
&&=-x^{-1}_2\delta\left(\frac{-x_0+x_1}{x_2}\right)\res_{y_1}
x^{-1}_1\delta\left(\frac{z-y_1}{-x_1}\right)\cdot \nno\\
&&\hspace{4em}\cdot \lambda(Y_1^*(e^{zL(-1)}v_2, x_{2})w_{(1)}\otimes Y_2(v_1,
y_1)w_{(2)})\nno\\
&&\quad +x^{-1}_2\delta\left(\frac{x_1-x_0}{x_2}\right)\res_{y_1}
x^{-1}_1\delta\left(\frac{z-y_1}{-x_1}\right)\cdot \nno\\
&&\hspace{4em}\cdot \lambda(w_{(1)}\otimes Y_2(v_2, -x_0+y_1)
Y_2(v_1, y_1)w_{(2)}).
\end{eqnarray}
Since
$$\res_{y_{2}}y_{2}^{-1}\delta\left(\frac{-x_{0}+y_{1}}{y_{2}}\right)=1,$$
the right-hand side of (10.17) can be written as
\begin{eqnarray}
&&-x^{-1}_2\delta\left(\frac{-x_0+x_1}{x_2}\right)\res_{y_1}
x^{-1}_1\delta\left(\frac{z-y_1}{-x_1}\right)\cdot \nno\\
&&\hspace{4em}\cdot \lambda(Y_1^*(e^{zL(-1)}v_2, x_{2})w_{(1)}\otimes Y_2(v_1,
y_1)w_{(2)})\nno\\
&&\quad +x^{-1}_2\delta\left(\frac{x_1-x_0}{x_2}\right)\res_{y_1}
x^{-1}_1\delta\left(\frac{z-y_1}{-x_1}\right)\res_{y_2}
y^{-1}_2\delta\left(\frac{-x_0+y_1}{y_2}\right)\cdot \nno\\
&&\hspace{4em}\cdot \lambda(w_{(1)}\otimes Y_2(v_2, -x_0+y_1)
Y_2(v_1, y_1)w_{(2)}).
\end{eqnarray}

Again by the properties of the formal $\delta$-function, (10.18) becomes
\begin{eqnarray}
&&x^{-1}_0\delta\left(\frac{x_2-x_1}{-x_0}\right)
\res_{y_1} x^{-1}_1\delta\left(\frac{z-y_1}{-x_1}\right)\cdot \nno\\
&&\hspace{4em}\cdot \lambda(Y_1^*(e^{zL(-1)}v_2, x_2)w_{(1)}\otimes Y_2(v_1,
y_1)w_{(2)})\nno\\
&&\quad +x^{-1}_2\delta\left(\frac{x_1-x_0}{x_2}\right)\res_{y_1}
x^{-1}_1\delta\left(\frac{z-y_1}{-x_1}\right)\res_{y_2}
y^{-1}_2\delta\left(\frac{-x_0+y_1}{y_2}\right)\cdot \nno\\
&&\hspace{4em}\cdot \lambda(w_{(1)}\otimes Y_2(v_2, y_2)Y_2(v_1, y_1)
w_{(2)})\nno\\
&&=x^{-1}_0\delta\left(\frac{x_2-x_1}{-x_0}\right)\res_{y_1}
x^{-1}_1\delta\left(\frac{z-y_1}{-x_1}\right)\cdot \nno\\
&&\hspace{4em}\cdot \lambda(Y_1^*(e^{zL(-1)}v_2, x_2)w_{(1)}\otimes Y_2(v_1,
y_1)w_{(2)})\nno\\
&&\quad +x^{-1}_2\delta\left(\frac{x_1-x_0}{x_2}\right)\res_{y_1}
\res_{y_2} x^{-1}_1\delta\left(\frac{z-y_1}{-x_1}\right)
y^{-1}_2\delta\left(\frac{-x_0+y_1}{y_2}\right)\cdot \nno\\
&&\hspace{4em}\cdot \lambda(w_{(1)}\otimes Y_2(v_2, y_2)Y_2(v_1, y_1)
w_{(2)})\nno\\
&&=x^{-1}_0\delta\left(\frac{x_2-x_1}{-x_0}\right)\res_{y_1}
x^{-1}_1\delta\left(\frac{z-y_1}{-x_1}\right)\cdot \nno\\
&&\hspace{4em}\cdot \lambda(Y_1^*(e^{zL(-1)}v_2, x_2)w_{(1)}\otimes Y_2(v_1,
y_1)w_{(2)})\nno\\
&&\quad -x^{-1}_2\delta\left(\frac{x_1-x_0}{x_2}\right)\res_{y_1} \res_{y_2}
\cdot\nno\\
&&\hspace{3em}\cdot
(x_2+x_0)^{-1}\delta\left(\frac{z-y_1}{-x_2-x_0}\right)
x^{-1}_0\delta\left(\frac{y_2-y_1}{-x_0}\right)\cdot \nno\\
&&\hspace{4em}\cdot \lambda(w_{(1)}\otimes Y_2(v_2, y_2)Y_2(v_1, y_1)
w_{(2)})\nno\\
&&=x^{-1}_0\delta\left(\frac{x_2-x_1}{-x_0}\right)\res_{y_1}
x^{-1}_1\delta\left(\frac{z-y_1}{-x_1}\right)\cdot \nno\\
&&\hspace{4em}\cdot \lambda(Y_1^*(e^{zL(-1)}v_2, x_2)w_{(1)}\otimes Y_2(v_1,
y_1)w_{(2)})\nno\\
&&\quad -x^{-1}_2\delta\left(\frac{x_1-x_0}{x_2}\right)\res_{y_1} \res_{y_2}
x^{-1}_2\delta\left(\frac{z-y_2}{-x_2}\right)x^{-1}_0\delta\left(\frac{y_2-
y_1}{-x_0}\right)\cdot \nno\\
&&\hspace{4em}\cdot \lambda(w_{(1)}\otimes Y_2(v_2, y_2)Y_2(v_1, y_1)w_{(2)}).
\end{eqnarray}
Substituting  (10.16)--(10.19) into (10.9) we obtain
\begin{eqnarray}
\lefteqn{\left(x^{-1}_0\delta\left(\frac{x_1-x_2}{x_0}\right)
Y'_{Q(z)}(v_1, x_1)
Y'_{Q(z)}(v_2,
x_2) \lambda\right)(w_{(1)}\otimes w_{(2)})}\nno\\
&&=x^{-1}_0\delta\left(\frac{x_1-x_2}{x_0}\right)\lambda(Y_1^*(e^{zL(-
1)}v_2, x_2)Y_1^*(e^{zL(-
1)}v_1, x_1)w_{(1)}\otimes w_{(2)})\nno\\
&&\quad -x^{-1}_0\delta\left(\frac{x_1-x_2}{x_0}\right)
\res_{y_2} x^{-1}_2\delta\left(\frac{z-y_2}{-x_2}\right)\cdot\nno\\
&&\hspace{6em}\cdot\lambda(Y_1^*(e^{zL(-1)}v_1, x_1)w_{(1)}\otimes Y_2(v_2,
x_2)w_{(2)})\nno\\
&&\quad -x^{-1}_0\delta\left(\frac{x_2-x_1}{-x_0}\right)\res_{y_1}
x^{-1}_1\delta\left(\frac{z-y_1}{-x_1}\right)\cdot\nno\\
&&\hspace{6em}\cdot\lambda(Y_1^*(e^{zL(-1)}v_2, x_2)w_{(1)}\otimes Y_2(v_1,
y_1)w_{(2)})\nno\\
&&\quad +x^{-1}_2\delta\left(\frac{x_1-x_0}{x_2}\right)\res_{y_1}
\res_{y_2} x^{-1}_2\delta\left(\frac{z-y_2}{-x_2}\right)
x^{-1}_0\delta\left(\frac{y_2-y_1}{-x_0}\right)\cdot\nno\\
&&\hspace{6em}\cdot\lambda(w_{(1)}\otimes Y_2(v_2, y_2)Y_2(v_1, y_1)w_{(2)}).
\end{eqnarray}

Now consider the result of the calculation {}from (10.16) to (10.19)
except for the last two steps in (10.19). Reversing
the subscripts 1 and 2 of the symbols $v$, $x$ and $y$ and replacing
$x_0$ by $-x_0$ in this result and then using a
property of the formal $\delta$-function, we have
\begin{eqnarray}
\lefteqn{x^{-1}_0\delta\left(\frac{x_2-x_1}{-x_0}\right)\res_{y_2}
x^{-1}_2\delta\left(\frac{z-y_2}{-x_2}\right)\cdot }\nno\\
&&\hspace{6em}\cdot (Y'_{Q(z)}(v_1, x_1) \lambda)
(w_{(1)}\otimes Y_2(v_2, y_2)w_{(2)})\nno\\
&& =x^{-1}_0\delta\left(\frac{x_1-x_2}{x_0}\right)\res_{y_2}
x^{-1}_2\delta\left(\frac{z-y_2}{-x_2}\right)\cdot \nno\\
&& \hspace{6em}\cdot \lambda(Y_1^*(e^{zL(-1)}v_1, x_1)w_{(1)}\otimes Y_2(v_2,
x_2)w_{(2)})\nno\\
&&\quad -x^{-1}_2\delta\left(\frac{x_1-x_0}{x_2}\right)\res_{y_1}\res_{y_2}
x^{-1}_2\delta\left(\frac{z-y_2}{-x_2}\right)
x^{-1}_0\delta\left(\frac{y_1-y_2}{x_0}\right)\cdot \nno\\
&&\hspace{6em}\cdot \lambda(w_{(1)}\otimes Y_2(v_1, y_1)Y_2(v_2, y_2)w_{(2)}).
\end{eqnarray}
{}From (10.5)--(10.9), again reversing the subscripts 1 and 2 of the symbols
$v$, $x$ and $y$
and replacing $x_0$ by $-x_0$, and (10.21), we have
 \begin{eqnarray}
\lefteqn{\left(-x^{-1}_0\delta\left(\frac{x_2-x_1}{-x_0}\right)Y'_{Q(z)}(v_2,
x_2)Y'_{Q(z)}(v_1, x_1)\cdot \lambda\right) (w_{(1)}\otimes w_{(2)})}\nno\\
&&=-x^{-1}_0\delta\left(\frac{x_2-x_1}{-
x_0}\right)\lambda(Y_1^*(e^{zL(-1)}v_1, x_1)Y_1^*(e^{zL(-
1)}v_2, x_2)w_{(1)}\otimes w_{(2)})\nno\\
&&\quad +x^{-1}_0\delta\left(\frac{x_2-x_1}{-x_0}\right)\res_{y_1}
x^{-1}_1\delta\left(\frac{z-y_1}{-x_1}\right)\cdot \nno\\
&&\hspace{6em}\cdot \lambda(Y_1^*(e^{zL(-1)}v_2,
x_2)w_{(1)}\otimes Y_2(v_1, y_1)w_{(2)})\nno\\
&&\quad +x^{-1}_0\delta\left(\frac{x_1-x_2}{x_0}\right)\res_{y_2}
x^{-1}_2\delta\left(\frac{z-y_2}{-x_2}\right)\cdot \nno\\
&&\hspace{6em}\cdot \lambda(Y_1^*(e^{zL(-1)}v_1, x_1)w_{(1)}\otimes Y_2(v_2,
x_2)w_{(2)})\nno\\
&&-x^{-1}_2\delta\left(\frac{x_1-x_0}{x_2}\right)\res_{y_1}\res_{y_2}
x^{-1}_2\delta\left(\frac{z-y_2}{-x_2}\right) x^{-1}_0\delta\left(\frac{y_1-
y_2}{x_0}\right)\cdot \nno\\
&&\hspace{6em}\cdot \lambda(w_{(1)}\otimes Y_2(v_1, y_1)Y_2(v_2, y_2)w_{(2)}).
\end{eqnarray}

The formulas (10.20) and (10.22) give:
\begin{eqnarray}
\lefteqn{\biggl(\biggl(x^{-1}_0\delta\left(\frac{x_1-x_2}{x_0}\right)
Y'_{Q(z)}(v_1, x_1)
Y'_{Q(z)}(v_2,
x_2)}\nno\\
&&\quad  -x^{-1}_0\delta\left(\frac{x_2-x_1}{-x_0}\right)
Y'_{Q(z)}(v_2, x_2)Y'_{Q(z)}(v_1,
x_1)\biggr) \lambda\biggr)(w_{(1)}\otimes w_{(2)})\nno\\
&&=\lambda\biggl(\biggl(x^{-1}_0\delta\left(\frac{x_1-x_2}{x_0}\right)
Y_1^*(e^{zL(-
1)}v_2, x_2)Y_1^*(e^{zL(-1)}v_1, x_1)\nno\\
&&\hspace{2em}-x^{-1}_0\delta\left(\frac{x_2-x_1}{-x_0}\right)
Y_1^*(e^{zL(-1)}v_1,
x_1)Y_1^*(e^{zL(-1)}v_2, x_2)\biggr)w_{(1)}\otimes w_{(2)}\biggr)\nno\\
&&\quad -x^{-1}_2\delta\left(\frac{x_1-x_0}{x_2}\right)\res_{y_1}\res_{y_2}
x^{-1}_2\delta\left(\frac{z-y_2}{-x_2}\right)\cdot \nno\\
&&\hspace{4em}\cdot \lambda\biggl(w_{(1)}\otimes \biggl(x^{-1}_0
\delta\left(\frac{y_1-y_2}{x_0}\right)
Y_2(v_1, y_1)Y_2(v_2, y_2)\nno\\
&&\hspace{2em}-x^{-1}_0\delta\left(\frac{y_2-y_1}{-x_0}\right)
Y_2(v_2, y_2)Y_2(v_1,
y_1)\biggr)w_{(2)}\biggr).
\end{eqnarray}
{}From the Jacobi identities for $Y_{1}^{*}$ and $Y_{2}$ and the
exponentiated {}form of the formula for the commutator of $L(-1)$ and
the vertex operators $Y(\cdot, x_{0})$ (cf. (10.7)), the right-hand
side of (10.23) is equal to
\begin{eqnarray}
&&x^{-1}_2\delta\left(\frac{x_1-x_0}{x_2}\right)\lambda(Y_1^*(Y(e^{zL(-
1)}v_1, x_0)e^{zL(-1)}v_2, x_2)w_{(1)}\otimes w_{(2)})\nno\\
&&\quad -x^{-1}_2\delta\left(\frac{x_1-x_0}{x_2}\right)\res_{y_1}\res_{y_2}
x^{-1}_2\delta\left(\frac{z-y_2}{-x_2}\right) y^{-1}_2\delta\left(\frac{y_1-
x_0}{y_2}\right)\cdot \nno\\
&&\hspace{6em}\cdot \lambda(w_{(1)}\otimes Y_2(Y(v_1, x_0)v_2, y_2)w_{(2)})
\nno\\
&&=x^{-1}_2\delta\left(\frac{x_1-x_0}{x_2}\right)\lambda(Y_1^*(e^{zL(-
1)}Y(v_1, x_0)v_2, x_2)w_{(1)}\otimes w_{(2)})\nno\\
&&\quad -x^{-1}_2\delta\left(\frac{x_1-x_0}{x_2}\right)\res_{y_1}\res_{y_2}
x^{-1}_2\delta\left(\frac{z-y_2}{-x_2}\right)
y^{-1}_2\delta\left(\frac{y_1-
x_0}{y_2}\right)\cdot \nno\\
&&\hspace{6em}\cdot \lambda(w_{(1)}\otimes Y_2(Y(v_1, x_0)v_2, y_2)w_{(2)}).
\end{eqnarray}
Using (10.8), evaluating
$\res_{y_{1}}$ and then using the definition of $Y'_{Q(z)}$ (recall (5.4)),
we finally see that the
right-hand side of (10.24) is equal to
\begin{eqnarray}
&&x^{-1}_2\delta\left(\frac{x_1-x_0}{x_2}\right)\lambda(Y_1^*(Y(v_1,
x_0)v_2, x_2+z)w_{(1)}\otimes w_{(2)})\nno\\
&&\quad -x^{-1}_2\delta\left(\frac{x_1-x_0}{x_2}\right)\res_{y_2}
x^{-1}_2\delta\left(\frac{z-y_2}{-x_2}\right)\cdot \nno\\
&&\hspace{6em}\cdot \lambda(w_{(1)}\otimes Y_2(Y(v_1, x_0)v_2, y_2)
w_{(2)})\nno\\
&&=x^{-1}_2\delta\left(\frac{x_1-x_0}{x_2}\right)(Y'_{Q(z)}(Y(v_1,
x_0)v_2, x_2) \lambda)(w_{(1)}\otimes w_{(2)}),
\end{eqnarray}
proving Theorem 6.1 in \cite{HL}.

\renewcommand{\theequation}{\thesection.\arabic{equation}}
\renewcommand{\therema}{\thesection.\arabic{rema}}
\setcounter{equation}{0}
\setcounter{rema}{0}

\section{Stability of certain subspaces of $(W_{1}\otimes W_{2})^{*}$
under the action of vertex operators}

We prove Proposition 6.2 in this section.

 Let $\lambda$ be an element of $(W_{1}\otimes W_{2})^{*}$ satisfying
the compatibility condition. We first want to prove that
each coefficient in $x$ of $Y'_{Q(z)}(u, x_0)Y'_{Q(z)}(v, x)\lambda$ is a
Laurent series involving only finitely many negative powers of $x_0$ and that
\begin{eqnarray}
\lefteqn{\tau_{Q(z)}\left(z^{-1}\delta\left(\frac{x_{1}-x_0}{z}\right)
Y_t(u, x_0)\right)
Y'_{Q(z)}(v, x)\lambda}\nno\\
&&=z^{-1}\delta\left(\frac{x_{1}-x_0}{z}\right)
Y'_{Q(z)}(u, x_0)Y'_{Q(z)}(v,
x) \lambda
\end{eqnarray}
for all $u, v\in V$.
Using the commutator formula for $Y'_{Q(z)}$, we have
\begin{eqnarray}
\lefteqn{Y'_{Q(z)}(u, x_0)Y'_{Q(z)}(v, x)\lambda=}\nno\\
&&=Y'_{Q(z)}(v, x)Y'_{Q(z)}(u, x_0)\lambda\nno\\
&&\quad -\res_{y}x^{-1}_0\delta\left(\frac{x-y}{x_0}\right)
Y'_{Q(z)}(Y(v, y)u, x_0)\lambda.
\end{eqnarray}
Each coefficient in $x$ of the right-hand side of
(11.2) is a
Laurent series involving only finitely many negative powers of $x_0$
 since $\lambda$
satisfies the lower truncation condition.  Thus the coefficients
in $x$ of $Y'_{Q(z)}(v, x)\lambda$ satisfy the lower truncation condition.

By (5.2) and (5.4), we have
\begin{eqnarray}
\lefteqn{\left(\tau_{Q(z)}\left(z^{-1}\delta\left(\frac{x_{1}-x_0}{z}\right)
Y_t(u, x_0)\right)
Y'_{Q(z)}(v, x) \lambda\right)(w_{(1)}\otimes w_{(2)})}\nno\\
&&=x^{-1}_0\delta\left(\frac{x_1-z}{x_0}\right)(Y'_{Q(z)}(v, x)
\lambda)(Y_1^*(u, x_1)w_{(1)}\otimes w_{(2)})\nno\\
&&\quad -x^{-1}_0\delta\left(\frac{z-x_1}{-x_0}\right)(Y'_{Q(z)}(v, x)
\lambda)(w_{(1)}\otimes Y_2(u, x_1)w_{(2)})\nno\\
&&=x^{-1}_0\delta\left(\frac{x_1-z}{x_0}\right)\biggl(\lambda(Y_1^*(v,
x+z)Y_1^*(u, x_1)w_{(1)}\otimes w_{(2)})\nno\\
&&\quad \quad -\res_{x_2}x^{-1}\delta\left(\frac{z-x_2}{-x}\right)
\lambda(Y_1^*(u,
x_1)w_{(1)}\otimes Y_2(v, x_2)w_{(2)})\biggr)\nno\\
&&\quad -x^{-1}_0\delta\left(\frac{z-x_1}{-x_0}\right)
\biggl(\lambda(Y_1^*(e^{zL(-
1)}v, x)w_{(1)}\otimes Y_2(u, x_1)w_{(2)})\nno\\
&&\quad \quad-\res_{x_2} x^{-1}\delta\left(\frac{z-x_2}{-x}\right)
\lambda(w_{(1)}\otimes
Y_2(v, x_2)Y_2(u, x_1)w_{(2)})\biggr).\;\;\;\;\;
\end{eqnarray}
Now the distributive law applies, giving us four terms. Inserting
$$\res_{x_{4}}x_{4}^{-1}\delta\left(\frac{x+z}{x_{4}}\right)=1$$
into the first of these terms and correspondingly replacing $x+z$ by $x_{4}$
in $Y^{*}_{1}(v, x+z)$, we can apply the commutator formula for $Y^{*}_{1}$
in the usual way. Also using the commutator formula for $Y_{2}$
and  properties of the $\delta$-function, we write the
 right-hand side of (11.3) as
\begin{eqnarray}
&&x^{-1}_0\delta\left(\frac{x_1-z}{x_0}\right)\lambda(Y_1^*(u, x_1)Y_1^*(v,
x+z)w_{(1)}\otimes w_{(2)})\nno\\
&&\quad -x^{-1}_0\delta\left(\frac{x_1-z}{x_0}\right)\res_{x_4}\res_{x_3}
x^{-1}_1\delta\left(\frac{x_{4}-x_3}{x_1}\right)
x^{-1}_4\delta\left(\frac{x+z}{x_4}\right)\cdot \nno\\
&&\hspace{6em}\cdot \lambda(Y_1^*(Y(v, x_3)u, x_1)w_{(1)}\otimes
w_{(2)})\nno\\
&&\quad -x^{-1}_0\delta\left(\frac{x_1-z}{x_0}\right)\res_{x_2}
x^{-1}\delta\left(\frac{z-x_2}{-x}\right)\lambda(Y_1^*(u, x_1)w_{(1)}\otimes
Y_2(v, x_2)w_{(2)})\nno\\
&&\quad -x^{-1}_0\delta\left(\frac{z-x_1}{-x_0}\right)\lambda(Y_1^*(e^{zL(-
1)}v, x)w_{(1)}\otimes Y_2(u, x_1)w_{(2)})\nno\\
&&\quad +x^{-1}_0\delta\left(\frac{z-x_1}{-x_0}\right)\res_{x_2}
x^{-1}\delta\left(\frac{z-x_2}{-x}\right)\lambda(w_{(1)}\otimes
Y_2(u, x_1)Y_2(v, x_2)w_{(2)})\nno\\
&&\quad +x^{-1}_0\delta\left(\frac{z-x_1}{-x_0}\right)\res_{x_2}
x^{-1}\delta\left(\frac{z-x_2}{-x}\right)\res_{x_3}
x^{-1}_1\delta\left(\frac{x_{2}-x_3}{x_1}\right)\cdot \nno\\
&&\hspace{6em}\cdot \lambda(w_{(1)}\otimes Y_2(Y(v, x_3)u, x_1)w_{(2)})\nno\\
&&=\biggl(x^{-1}_0\delta\left(\frac{x_1-z}{x_0}\right)\lambda(Y_1^*(u,
x_1)Y_1^*(e^{zL(-1)}v, x)w_{(1)}\otimes w_{(2)})\nno\\
&&\quad \quad-x^{-1}_0\delta\left(\frac{z-x_1}{-x_0}\right)
\lambda(Y_1^*(e^{zL(-
1)}v, x)w_{(1)}\otimes Y_2(u, x_1)w_{(2)})\biggr)\nno\\
&&\quad -\res_{x_2} x^{-1}\delta\left(\frac{z-x_2}{-x}\right)
\biggl(x^{-1}_0\delta\left(\frac{x_1-z}{x_0}\right)\cdot\nno\\
&&\hspace{6em}\cdot\lambda(Y_1^*(u,
x_1)w_{(1)}\otimes Y_2(v, x_2)w_{(2)})\nno\\
&&\quad \quad -x^{-1}_0\delta\left(\frac{z-x_1}{-x_0}\right)
\lambda(w_{(1)}\otimes
Y_2(u, x_1)Y_2(v, x_2)w_{(2)})\biggr)\nno\\
&&\quad -\biggl(\!\! \res_{x_4}\res_{x_3}(x_0+z)^{-1}
\delta\left(\frac{x+z-x_3}{x_0+
z}\right) \cdot\nno\\
&&\hspace{5em}\cdot x^{-1}_4\delta\left(\frac{x+z}{x_4}\right)
x^{-1}_0\delta\left(\frac{x_1-z}{x_0}\right)\cdot \nno\\
&&\hspace{6em}\cdot \lambda(Y_1^*(Y(v, x_3)u, x_1)w_{(1)}\otimes w_{(2)})\nno\\
&&\quad \quad-\res_{x_2}\res_{x_3}x^{-1}\delta\left(\frac{z-(x_1+x_3)}{-
x}\right) x^{-1}_2\delta\left(\frac{x_1+x_3}{x_2}\right)\cdot \nno\\
&&\hspace{6em}\cdot x^{-1}_0
\delta\left(\frac{z-
x_1}{-x_0}\right)\lambda(w_{(1)}\otimes Y_2(Y(v, x_3)u, x_1)w_{(2)})\biggr).
\end{eqnarray}

By (5.2) and  properties of formal
$\delta$-function, the right-hand side of (11.4) is equal to
\begin{eqnarray}
&&\left(\tau_{Q(z)}\left(z^{-1}\delta\left(\frac{x_{1}-x_0}{z}\right)Y_t(u,
x_0)\right)\lambda\right)(Y_1^*(e^{zL(-1)}v, x)w_{(1)}\otimes w_{(2)})\nno\\
&&\quad -\res_{x_2} x^{-1}\delta\left(\frac{z-x_2}{-x}\right)\cdot \nno\\
&&\hspace{4em}\cdot
\left(\tau_{Q(z)}\left(z^{-1}\delta\left(\frac{x_{1}-x_0}{z}\right)
Y_t(u, x_0)\right)\lambda\right)(w_{(1)}\otimes Y_2(v, x_2)w_{(2)})\nno\\
&&\quad -\biggl(\res_{x_3}x^{-1}_0
\delta\left(\frac{x-x_3}{x_0}\right)\cdot\nno\\
&&\hspace{4em}\cdot x^{-1}_0\delta\left(\frac{x_1-z}{x_0}\right)
\lambda(Y_1^*(Y(v, x_3)u,
x_1)w_{(1)}\otimes w_{(2)})\nno\\
&&\quad -\res_{x_3}x^{-1}\delta
\left(\frac{x_0+x_{3}}{x}\right)\cdot \nno\\
&&\hspace{4em}\cdot x^{-1}_0\delta\left(\frac{z-x_1}{-x_0}\right)
\lambda(w_{(1)}\otimes Y_2(Y(v, x_3)u, x_1)w_{(2)})\biggr).
\end{eqnarray}
Using (5.2) and (5.4), we find that the right-hand side of
(11.5) becomes
\begin{eqnarray}
&&\left(Y'_{Q(z)}(v, x)\tau_{Q(z)}\left(z^{-1}
\delta\left(\frac{x_{1}-x_0}{z}\right)
Y_t(u, x_0) \right)\lambda\right)(w_{(1)}\otimes w_{(2)})\nno\\
&&\quad -\biggl(\!\! \res_{x_3} x^{-1}_0
\delta\left(\frac{x-x_3}{x_0}\right)\cdot \nno\\
&&\hspace{3em}\cdot \tau_{Q(z)}\biggl(z^{-1}\delta\left(\frac{x_{1}-x_0}{z}
\right)
Y_t(Y(v, x_3)u,
x_0)\biggr) \lambda\biggr)(w_{(1)}\otimes w_{(2)}).\;\;\;\;\;
\end{eqnarray}
By the compatibility condition for $\lambda$ and the commutator formula for
$Y'_{Q(z)}$, the right-hand side of (11.6) is
equal to
\begin{eqnarray}
&&z^{-1}\delta\left(\frac{x_{1}-x_0}{z}\right)
(Y'_{Q(z)}(v, x)Y'_{Q(z)}(u,
x_0) \lambda)(w_{(1)}\otimes w_{(2)})\nno\\
&&\quad -z^{-1}\delta\left(\frac{x_{1}-x_0}{z}\right)\biggl(\!\! \res_{x_3}
x^{-1}_0\delta\left(\frac{x-x_3}{x_0}\right)\cdot \nno\\
&&\hspace{6em}\cdot Y'_{Q(z)}(Y(v, x_3)u, x_0)
\lambda\biggr)(w_{(1)}\otimes w_{(2)})\nno\\
&&= z^{-1}\delta\left(\frac{x_{1}-x_0}{z}\right)
\biggl(\biggl(Y'_{Q(z)}(v, x)Y'_{Q(z)}(u, x_0)\nno\\
&&\quad -\res_{x_3} x^{-1}_0\delta\left(\frac{x-x_3}{x_0}\right)
Y'_{Q(z)}(Y(v, x_3)u,
x_0)\biggr) \lambda\biggr)(w_{(1)}\otimes w_{(2)})\nno\\
&&=z^{-1}\delta\left(\frac{x_{1}-x_0}{z}\right)
(Y'_{Q(z)}(u, x_0)Y'_{Q(z)}(v,
x) \lambda)(w_{(1)}\otimes w_{(2)}).
\end{eqnarray}
This proves (11.1) and hence the first part of  Proposition 6.2.

Finally, suppose that $\lambda\in (W_{1}\otimes W_{2})^{*}$ satisfies
the local grading-restriction condition.  We want to prove that for
any $v\in V$, the components of $$Y'_{Q(z)}(v, x)\lambda =\sum_{n\in
{\Bbb Z}}\tau_{Q(z)}(v\otimes t^{n})\lambda x^{-n-1}$$ also satisfy the
local grading-restriction condition. {}From (9.6), for $n\in {\Bbb Z}$
the coefficient $\tau_{Q(z)}(v\otimes t^{n})$ of $x^{-n-1}$ in
$Y'_{Q(z)}(v, x)$ has weight $\mbox{\rm wt}\ v -n-1$, that is,
$\tau_{Q(z)}(v\otimes t^{n})$ maps a weight vector of weight $h$ to a
weight vector of weight $\mbox{\rm wt}\ v -n-1+h$.  Thus the
components of $Y'_{Q(z)}(v, x)\lambda$ are sums of weight vectors
since $\lambda$ is a sum of weight vectors. Now note that for any
$v\in V$ and $n\in {\Bbb Z}$, the (graded) subspace
$W_{\tau_{Q(z)}(v\otimes t^{n})\lambda}$ (recall the notation in part
(b) of the local grading-restriction condition) is contained in the
subspace $W_{\lambda}$ and since $W_{\lambda}$ satisfies (6.2) and
(6.3), so must $W_{\tau_{Q(z)}(v\otimes t^{n})\lambda}$.  Thus
$\tau_{Q(z)}(v\otimes t^{n})\lambda$ satisfies the local
grading-restriction condition for all $v\in V$ and $n\in {\Bbb Z}$,
and the second and remaining part of Proposition 6.2 is proved.

 {\small \sc Department of Mathematics, University of Pennsylvania,
Philadelphia, PA 19104}

 {\it Current address:} Department of Mathematics, Rutgers University,
New Brunswick, NJ 08903

{\em E-mail address}: yzhuang@math.rutgers.edu

\vskip 1em

{\small \sc Department of Mathematics, Rutgers University,
New Brunswick, NJ 08903}

{\em E-mail address}: lepowsky@math.rutgers.edu

\end{document}